\RequirePackage{fix-cm}
\documentclass[twocolumn,epjc3]{svjour3}  
\smartqed  
\RequirePackage{graphicx}
\RequirePackage{amsmath}
\RequirePackage{mathptmx}      
\RequirePackage{latexsym}
\RequirePackage{cuted}
\RequirePackage[numbers,sort&compress]{natbib}
\RequirePackage[colorlinks,citecolor=blue,urlcolor=blue,linkcolor=blue]{hyperref}
\newcommand{\p}{{\mathrm {p}}}
\newcommand{\DM}{\hbox{\tiny{DM}}}
\newcommand{\B}{\hbox{\tiny{B}}}
\newcommand{\ini}{\scriptsize{\hbox{ini}}}
\newcommand{\peak}{\hbox{\scriptsize{peak}}}
\journalname{Eur. Phys. J. C}
\begin{document}

\title{Non--comoving baryons and cold dark matter in cosmic voids
}

\titlerunning{Non--comoving baryons and cold dark matter in cosmic voids}        

\author{Ismael Delgado Gaspar\thanksref{e1,addr1}
        \and
        Juan Carlos Hidalgo\thanksref{e2,addr1} 
        \and
        Roberto A. Sussman\thanksref{e3,addr2} 
}

\thankstext{e1}{e--mail: ismaeldg@icf.unam.mx}
\thankstext{e2}{e--mail: hidalgo@icf.unam.mx}
\thankstext{e3}{e--mail: sussman@nucleares.unam.mx}


\institute{Instituto de Ciencias F\'isicas, Universidad Nacional Aut\'onoma de M\'exico, A.P. 48--3, 62251
Cuernavaca, Morelos, M\'exico. \label{addr1}
           \and
           Instituto de Ciencias Nucleares, Universidad Nacional Aut\'onoma de M\'exico, A. P. 70--543,
04510 Ciudad de M\'exico, M\'exico. \label{addr2}
}

\date{Received: date / Accepted: date}

\maketitle

\begin{abstract}
We examine the fully relativistic evolution of cosmic voids constituted by baryons and cold dark matter (CDM), represented by two non--comoving dust sources in a $\Lambda$CDM background. For this purpose, we consider numerical solutions of Einstein's field equations in a fluid--flow representation adapted to spherical symmetry and multiple components.  
We present a simple example that explores the frame--dependence of the local expansion and the Hubble flow for this mixture of two dusts, revealing that the relative velocity between the sources yields a significantly different evolution in comparison with that of the two sources in a common 4--velocity (which reduces to a Lema\^\i tre--Tolman--Bondi model). In particular, significant modifications arise for the density contrast depth and void size, as well as in the amplitude of the surrounding over--densities. 
We show that an adequate model of a frame--dependent evolution that incorporates initial conditions from peculiar velocities and large--scale 
density contrast observations may contribute to understand the discrepancy between the local value of $H_0$ and that inferred from the CMB. 

\end{abstract}

\section{Introduction}\label{SecIntro}
The generic term ``Cosmic Voids'' denotes $\sim10-120 $ Mpc sized round shaped low density regions surrounded by overdense filaments and walls, all of which conform the ``cosmic web'' of large--scale structure (baryons and CDM) revealed by observations and N--body simulations \cite{libeskind2018tracing}.
%
There is a large body of literature on cosmic voids, from seminal early work \cite{peebles2001void,friedmann2001model} to recent extensive reviews~\cite{van2011cosmic,van2014voids,sutter2014life,hamaus2014modeling} and detailed catalogues (see a summary in~\cite{Catalogos}).  As revealed by these reviews and references therein: (i)  cosmic voids enclose only 15\% of cosmic matter--energy (within the $\Lambda$CDM paradigm) but constitute about 77\% of cosmic volume; (ii) they  form from early negative density contrast perturbations; (iii) they roughly keep their rounded shape and (iv) their dynamics is relatively insensitive to considerations from baryon physics. This relatively simple and pristine dynamics renders them as ideal structure systems to improve the theoretical modeling of generic cosmological  
observations~\cite{xie2014local,hamaus2016constraints,nadathur2016testing}, and to assess several open problems in cosmology: the nature of dark 
matter and dark 
energy~\cite{lee2009constraining,alexander2009local,biswas2010voids,de2011effects,bos2012darkness,roukema2013dark,sutter2014observability,pisani2015counting,adermann2018cosmic,pollina2018relative}, 
redshift space 
distortions~\cite{hamaus2015probing,achitouv2017improved,nan2018gravitational,maeda2011dynamics}, 
Cosmic Microwave Background (CMB) properties~\cite{cai2014possible,inoue2006local,nadathur2011reconciling,chantavat2016cosmological,cai2016lensing}, Baryonic Acoustic Oscillations (BAO)~\cite{zhao2018improving}, alternative gravity 
theories~\cite{PhysRevLett.117.091302,zivick2015using,spolyar2013topology,voivodic2017modeling,sahlen2018cluster,clampitt2013voids}, local group kinematics and peculiar velocity fields~\cite{tully2006our,davies2008local,tully2008our,lavaux2010dynamics,nasonova2011kinematics,lares2017voids,ahn2018formation}, as well as theoretical issues such as gravitational entropy~\cite{sussman2015gravitational,mishra2014thermodynamics}.       

The usual approach to study the nature and dynamics of cosmic voids is through Newtonian gravity~\cite{demchenko2016testing} 
(see reviews~ \cite{van2011cosmic,van2014voids,sutter2014life,hamaus2014modeling} and references cited therein for examples of studies 
based on analytic work, perturbations and N--body simulations). 
These studies have put forward various forms of ``universal'' density profiles~\cite{hamaus2014universal,ricciardelli2014universality} that fit 
observations and catalogues. However, given the fact that cosmic voids are approximately spherical structures that tend to 
become more spherical as they evolve (see \cite{Icke1984} for the first proof of this fact known as the ``\textit{bubble theorem}''\footnote{
The ``Bubble Theorem'' states that an isolated underdensity (void) tends to evolve into a spherical shape, explicitly: ``\textit{as the void becomes bigger, its 
asphericities will tend to disappear}'' \cite{Icke1984}.
}
and also~\cite{sato1983expansion,van2011cosmic,van2014voids,sutter2014life,hamaus2014modeling,wojtak2016voids,2018VoidNewtonSS} for further discussions and comparison with N--body simulations), 
it is also feasible to study them by means of spherically symmetric, exact and numerical solutions of Einstein's equations. As examples of analytic and semi--analytic general relativistic studies, there are many based on Lema\^\i tre--Tolman--Bondi (LTB) dust models~\cite{bolejko2005formation,bolejko2006radiation,BKHC2009,iribarrem2014relativistic}, 
or the more general non--spherical (but quasi--spherical) Szekeres models~\cite{BoSu2011,multi,paperJCAP,CPTSzek2017}. 
While numerical relativity techniques have already been applied in a cosmological context beyond spherical symmetry~\cite{mewes2018numerical,macpherson2017inhomogeneous,macpherson2018trouble,bentivegna2016effects,daverio2017numerical,bolejko2017relativistic}, 
most relativistic numerical studies on cosmic voids still rely on metric--based techniques involving the Misner--Sharp mass function, and thus their validity is restricted to spherical symmetry~\cite{torres2014cosmological,lasky2007spherically,lasky2006generalized,lim2013spherically}. 
In particular, four relevant studies that specifically examine general relativistic void dynamics for spherical symmetry preceding our study are~\cite{kim2018alternative,tsizh2016evolution,novosyadlyj2016evolution,Marra:2011zp}.   

In the present article we examine the fully general relativistic evolution of a spherically symmetric cosmic void, assuming as matter source a mixture of non--interactive baryons and CDM species, each evolving along a different 4--velocity.  Specifically, we consider the evolution of a generic cosmic void suitably embedded in a $\Lambda$CDM background. Since CDM is the dominant clustering source, 
we assume a frame in which its 4--velocity is comoving, whereas the baryons evolve along a non--comoving 4--velocity that defines a non--trivial 
field of spacelike relative velocities with respect to the CDM frame. Consequently the 4--velocities of the two dust sources are related by a boost, and the energy--momentum tensor of the mixture (as described in the comoving frame) no longer has the form of a perfect fluid, but that of a complicated ``fluid''  energy--momentum tensor that contains effective pressures and energy flux terms associated with the relative velocity field.   

In order to deal with the general relativistic dynamics for this energy--momentum tensor, we do not resort to the traditional metric based methods of ~\cite{kim2018alternative,tsizh2016evolution,novosyadlyj2016evolution,Marra:2011zp}, but consider the system of first order partial differential equations (PDEs) provided by the ``fluid--flow'' (or ``1+3'') representation of Einstein's equations~\cite{ellis1999cosmological, Tsagas:2007yx,ellis2012relativistic} in terms of evolution equations and constraints for the covariant quantities associated with the CDM comoving 4--velocity. 
These covariant variables are 
(i) kinematic: expansion scalar, 4--acceleration, shear tensor and relative velocity, all computed from the CDM frame; 
(ii) source terms: the total energy density, pressure and energy flux that arise from the relative velocity and 
(iii)  the electric Weyl tensor.  These equations (evolutions and constraints) must be supplemented by spacelike constraints.

The plan of the paper is as follows. In Section \ref{SecMixtureMultiFluidsSS} , we introduce a model for the evolution of a generic 
mixture of fluids in a spherically symmetric spacetime.  This model is specialized to the case of two non--interacting dust--like fluids, namely 
CDM and baryons, in Section \ref{Sec:VoidFormation}. 
We examine the numerical solutions of the resulting system of partial differential equations (PDEs) in a void formation scenario. Through representative numerical examples, we look at the influence of the relative baryon--CDM velocity on the 
evolution and present--day final structure. 
Our results are summarized and discussed in Section \ref{Sec:Disc-Remarks}. 
Finally, we have included three appendices that complement the main text. 
 \ref{App1plus3} provides the general tensorial evolution equations of the $1+3$ description; while in \ref{App:Dimensionless_system} and \ref{App:LTB_limit}, we present the dimensionless baryon--CDM system of PDEs and show its (single--fluid) LTB limit.


\section{A mixture of multiple fluids in spherical symmetry}\label{SecMixtureMultiFluidsSS}

A spherically symmetric spacetime is characterized by the line element,
\begin{eqnarray}
ds^2&=&g_{\mu\nu} dx^\mu dx^\nu \nonumber
\\
&=&-N^2 dt^2 + B^2 dr^2 +Y^2\left(d\theta^2+\sin^2(\theta) d\phi^2\right),\label{SSg_ab}
\end{eqnarray}
where the metric coefficients $N$, $B$ and $Y$ are functions of the radial and time coordinates. 
Notice that this metric contains as a particular case the Robertson--Walker line element, a solution recovered in our examples at scales larger than $\sim100$ Mpc at $z=0$.

Regarding the evolution of a mixture of non--comoving fluids, the fluid elements 
of each species will evidently present its own 4--velocity. 
In absence of specific criteria, a useful choice of ``reference frame'' is the one of the dominant species\footnote{This choice, however, is arbitrary and 
one could also choose the fundamental observers 
moving with the baryon component or even  a frame in which the total momentum flux density vanishes $q^\mu=0$.}. 
Hence the family of fundamental observers will evolve with 
comoving 4--velocity,  
\begin{equation}
    u^\mu = \frac{1}{N} \delta^\mu_{~t},
\end{equation}

\noindent which defines the convective fluid--flow (or time derivative) and space--like gradients (orthogonal to $u^\mu$):
\begin{equation}\label{eqn:convec-spacegrad}
\dot{X}=u^\mu \nabla_\mu X=\frac{X_{,t}}{N} \quad \hbox{and} \quad  \tilde{\nabla}_\mu X =h^\nu_{~\mu} \nabla_{\nu}X.
\end{equation}
The kinematic parameters associated with a spherically symmetric fluid as measured by 
the fundamental observers are the expansion scalar, 4--acceleration and shear tensor (the vorticity vanishes identically): 
\begin{subequations}
\begin{eqnarray}
H&=&\frac{\Theta}{3}=\frac13 \tilde\nabla_\mu u^\mu , 
\\
\dot{u}_\mu&=&u^\nu\nabla_\nu u_\mu, \label{acc_def}
\\
\sigma_{\mu\nu}&=&\tilde\nabla_{(\mu}u_{\nu)}-\dot{u}_\mu u_\nu-Hh_{\mu\nu}. \label{Shear_def}
\end{eqnarray}
\end{subequations}
The Einstein field equations can be recast~\cite{ellis1999cosmological, Tsagas:2007yx,ellis2012relativistic} as a first order system of 
``$1+3$'' evolution and constraint equations involving these kinematic parameters, together with the energy density $\rho$, isotropic $p$ and anisotropic pressure $\pi_{\mu\nu}$ and energy flux $q_\mu$ of the source given by the energy--momentum tensor, as well as the electric Weyl tensor $E_{\mu\nu}=C_{\mu\alpha\nu\beta} \, u^{\alpha}u^{\beta}$ (the magnetic Weyl tensor identically vanishes). This system (displayed in \ref{App1plus3}) is ideal for a numerical treatment in which all variables have a clear physical and geometric meaning. Since it is based on a 4--velocity flow it is fully covariant, and thus it is readily applicable to spacetimes that are not spherically symmetric.  

For the spherically symmetric metric (\ref{SSg_ab}) we have 
\begin{equation}H = \frac13\left(\frac{2\dot Y}{Y}+\frac{\dot B}{B}\right),\quad \dot u_\mu=\tilde\nabla_\mu\left(\ln N\right)= A\delta^r_\mu,\quad A\equiv \frac{N_{,r}}{N},\label{Hdotu}\end{equation}
and the space--like symmetric trace--free tensors $\sigma^\mu_{~\nu}$ and $E^\mu_{~\nu}$
can be written as
\begin{equation}\label{eqn:sigma-EeigenV}
\sigma^\mu_{~\nu}=\Sigma ~e^\mu_{~\nu}, \qquad\qquad E^\mu_{~\nu}=W ~e^\mu_{~\nu}.
\end{equation}
Here $W$ and $\Sigma$ are scalar functions,
\begin{equation}\label{eqn:sigma-EeigenV}
\Sigma=\frac{1}{3} \left(\frac{\dot{Y}}{Y}-\frac{\dot{B}}{B}\right) , \qquad W =-\Psi_2,
\end{equation}
with $\Psi_2$ the conformal invariant of Petrov type D spacetimes, and
$e^\mu_{~\nu} =h^\mu_{~\nu}-3n^\mu n_{\nu} =\mathrm{Diag}\left[0,-2,1,1\right]$ is the tensor basis that serves as eigenframe for 
spacelike symmetric trace free tensors in Petrov type D spacetimes, with $h_{\mu \nu}=g_{\mu \nu}+u_\mu u_\nu$ and 
$n_\mu=\sqrt{g_{rr}} \, \delta^r_{~\mu}$ the projection tensor and  a spacelike normal vector tangent to the orbits of SO$(3)$
 (note that  $\dot{e}^\mu_{~\nu} =0$). 

The 4--velocity of the other non--comoving components are related to $u^\mu$ via the relative velocity measured by the fundamental observers $v_{(i)}^\mu$, defined such that $v_{(i)}^\mu u_{\mu}=0$. Then, the 4--velocity of the i--th fluid reads,
\begin{equation}
u^\mu_{(i)}=\gamma_{(i)} \left(u^\mu+v_{(i)}^\mu\right), \quad \hbox{with} \quad \gamma_{(i)} =\left(1-v_{(i)}^2\right)^{-\frac{1}{2}},
\label{4velocityUi}
\end{equation}
where ``$i$'' labels the components and $v_{(i)}^2=g_{\mu\nu}v_{(i)}^\mu v_{(i)}^\nu$. 

The total energy--momentum tensor is made up of all the contributions from the different species, 
and in general it will no longer be the energy--momentum tensor of a perfect fluid, but
%
\begin{equation}
T^{\mu\nu}=\sum\limits_{i}T^{\mu\nu}_{(i)}=\rho\, u^\mu u^\nu + p \, h^{\mu\nu} + 2 q^{(\mu} u^{\nu)} + \pi^{\mu\nu}, \label{Tequal_Sum_Ti}
\end{equation}
where $\rho$, $p$, $\pi^{\mu\nu}$ and $q^{\mu}$ are the energy density, isotropic and anisotropic pressures\footnote{
In this setup, the cosmological constant is implicitly considered by the substitution $\rho \rightarrow \rho + \Lambda$ and $\p\rightarrow \p-\Lambda$.
}, and the energy 
flow measured by the fundamental observers along $u^\mu$. These components are determined by projecting the energy--momentum 
tensor parallel and orthogonal to
$u^\mu$ \cite{maartens1999cosmic,ellis2012relativistic}:
%
%
\begin{subequations}\label{TprojectionIyII}
\begin{eqnarray}
\rho&=&T^{\mu\nu} u_\mu u_\nu,   \;\;\; q^\mu=-T^{\mu\nu} u_\nu - \rho u^{\mu},  \;\;\;  p=\frac{1}{3} T^{\mu \nu} h_{\mu \nu},\nonumber\\\label{TprojectionI}
\\
\pi_{\mu \nu}&=&T_{\langle\mu \nu\rangle}=\left[h_{\,(\mu}^{\eta}h_{\, \nu)}^{\upsilon}-\frac{1}{3} h_{\mu \nu} h^{\eta \upsilon}\right] T_{\eta \upsilon}.\label{TprojectionII}
\end{eqnarray}
\end{subequations}
Although the total energy--momentum tensor is always conserved, the energy--momentum tensors of the individual components are not necessarily 
conserved. If there are 
non--gravi\-\ tational interactions 
between them, they satisfy $\nabla_\nu T^{\mu\nu}_{(i)}=J^\mu_{(i)}$, 
where $J_{(i)}$ is the rate of energy and  momentum densities transfer between species $i$ 
as measured in the $u^\mu$--frame. In absence of non--gravitational interaction these energy--momentum tensors are separately conserved: $
J^\mu_{(i)}=0$ for all $i$.
\subsection{A mixture of non--interacting perfect fluids}

We now focus on the case of a mixture of non--interacting fluids, each one a perfect fluid with a suitable equation of state in its intrinsic frame (denoted with ${}^*$):
\begin{equation}
p_{(i)}^\ast=w_{(i)} \,\rho_{(i)}^\ast, \qquad \hbox{in general} \qquad w_{(i)}=w_{(i)}(t,r).
\end{equation}
In this way, the total energy--momentum tensor follows from adding up the corresponding 
tensors of the dynamically significant species as seen from the $u^\mu$ frame~ (eqs. (\ref{Tequal_Sum_Ti}) and (\ref{TprojectionIyII})).  

Explicitly, if we choose the fundamental observers those along $u_{(0)}^\mu$, then the contribution to the total energy--momentum tensor 
of the ``$0$'' component reads
\begin{equation}
T_0^{\mu \nu}=\rho_0^\ast u^\mu u^\nu + p_0^\ast \, h^{\mu\nu}.\label{T_0}
\end{equation}
On the other hand, the energy--momentum tensor of the i--th component comoving with velocity $u_{(i)}^\mu$ (see eq.~(\ref{4velocityUi})) takes the form \begin{equation}
T_{(i)}^{\mu \nu}=\rho_{(i)} u^\mu u^\nu + p_{(i)} h^{\mu\nu} + 2 q_{(i)} ^{(\mu} u^{\nu)} + \pi_{(i)}^{\mu\nu}, \label{T_i}
\end{equation}
with the dynamical quantities given by~\cite{maartens1999cosmic,ellis2012relativistic}:
\begin{subequations}\label{subeqnTotalTherm}
\begin{eqnarray}
\rho&=&\sum\limits_{i}\rho_{(i)},\quad \rho_{(i)}=\gamma_{(i)}^2 (1+ w_{(i)} v_{(i)}^2 ) \rho^\ast_{(i)},
\label{RhoPerFluid_i}
\\
p&=&\sum\limits_{i} p_{(i)},\;\;\;\, p_{(i)} = \left[ w_{(i)} + \frac{1}{3} \gamma_{(i)}^2  v_{(i)}^2 (1+ w_{(i)}) \right] \rho^\ast_{(i)}, 
\label{PPerFluid_i}
\\
q^\mu &=& \sum\limits_{i}q^\mu_{(i)},\quad q^\mu_{(i)}= \gamma_{(i)}^2 (1+w_{(i)}) \rho^\ast_{(i)} v_{(i)}^\mu,
\label{qPerFluid_i}
\\
\pi^{\mu\nu} &=& \sum\limits_{i}\pi^{\mu\nu}_{(i)},\quad \pi^{\mu\nu}_{(i)}=\gamma_{(i)}^2 (1+w_{(i)}) \rho^\ast_{(i)} v_{(i)}^{\langle\mu} v_{(i)}^{\nu\rangle},
\label{AniPPerFluid_i}
\end{eqnarray}
\end{subequations}
where for spherical symmetry spacetimes the anisotropy tensor can be written as $\pi^{\mu}_{(i) \, \nu}=\Pi_{(i)} e^{\mu}_{~\nu}$, with $\Pi_{(i)}$
to be determined from Eq.~\eqref{AniPPerFluid_i}.

The dynamics of this fluid mixture can be determined from the first order ``$1+3$'' fluid flow  representation of Einstein's field equations given in \ref{App1plus3}, by direct substitution of $\rho,\,p,\,\pi_{\mu\nu},\,q_\mu$ by (\ref{RhoPerFluid_i})--(\ref{AniPPerFluid_i}), with
\begin{equation} q_{(i)\mu} = Q_{(i)}\delta_{~\mu}^r,\qquad v_{(i)\mu} =V_{(i)}\delta_{~\mu}^r,\label{qu_def}\end{equation}
and $\dot u_\mu,\,\sigma_{\mu\nu},\,E_{\mu\nu}$ obtained from (\ref{Hdotu}) and (\ref{eqn:sigma-EeigenV}). For spherical symmetry this system can be further simplified and complemented by evolution equations for the metric functions $Y,\,B$ and $\chi\equiv Y'$ that can be obtained from (\ref{Hdotu}) and (\ref{eqn:sigma-EeigenV}). Since each one of the proper tensors $\dot u_\mu,\,\sigma_{\mu\nu},\,E_{\mu\nu}$ is fully determined by a single scalar function $A,\,\Sigma,\,W$, the tensorial system in \ref{App1plus3} becomes a dynamical system involving only scalar functions:
\begin{subequations}\label{1plus3EFE}

\begin{eqnarray}
\dot Y &=& Y\left(H+\Sigma\right),
\\
\dot{\chi} &=&A \left(\Sigma+H\right) Y-2\chi \Sigma +\chi H+\frac{1}{2}\kappa Q Y,
\\
\dot B &=& B \left(H-2 \Sigma\right),
\\
\dot H&=& - H^2 -2\Sigma^2 - \frac{1}{6}\kappa \left(\rho+3p\right) +\frac{1}{3}\frac{A^2}{B^2} +\frac{1}{3}\frac{A_{,r}}{B^2}
\nonumber
\\
&{}&\qquad\qquad\qquad\qquad-\frac{1}{3} \frac{A B_{,r}}{B^3}+ \frac{2}{3} \frac{A  \chi }{ B^2 Y} , 
\\
\dot\Sigma&=&\Sigma^2- 2 H \Sigma+\frac{1}{2} \kappa \Pi - W-\frac{1}{3}\frac{A^2}{B^2}+ \frac{1}{3}\frac{A \chi}{B^2 Y} 
\nonumber
\\
&{}&\qquad\qquad\qquad\qquad+\frac{1}{3} \frac{A B_{,r}}{B^3}-\frac{1}{3} \frac{A _{,r}}{B^2},\label{dotSig}\\
\dot W&+& \frac{1}{2} \kappa \dot\Pi=-3 \left(H +\Sigma\right)W -\frac{1}{2} \kappa \left( \rho+p\right) \Sigma
\nonumber
\\
&{-}&\frac{1}{2} \kappa \left(H-\Sigma\right) \Pi
- \kappa \frac{Q B_{,r}}{6 B^3}+\kappa \frac{Q_{,r}}{6 B^2}-
\kappa\frac{Q \chi}{6 B^2 Y}+\kappa \frac{A Q}{3 B^2},\nonumber\\\label{dotW}
\end{eqnarray}
\end{subequations}
where $A$ is defined in (\ref{Hdotu}), $Q=\sum\limits_i Q_{(i)}$ and $\Pi=\sum\limits_i \Pi_{(i)}$.  This system must be complemented by the following constraints:
\begin{subequations}\label{Subeqs:ConsWHN}
\begin{eqnarray}
W &=& -\frac{1}{6}\kappa\rho-\frac{\kappa}{2}\Pi+\frac{M}{Y^3},\label{WMconstr}
\\
N_{,r}&=&A N,\label{eqn:NrAN}
\\
H^2&=&\frac{1}{3}\kappa \rho-\mathcal{K}+\Sigma^2.
\end{eqnarray}
\end{subequations}
Here $\mathcal{K}$ is the spatial curvature: 
\begin{equation}\mathcal{K}\equiv \frac{{}^{(3)}\mathcal{R}}{6}=\frac{(K Y)_{,r}}{3 Y^2 \chi}, \quad \hbox{with} \quad K=1-\frac{\chi^2}{B^2},\label{KurvEqn}\end{equation}
and $M$ is the Misner--Sharp function
\begin{equation}M = \frac{Y}{2}\left[\dot Y^2-\frac{\chi^2}{B^2}+1\right]=
\frac{Y}{2}\left[Y^2(H+\Sigma)^2-\frac{\chi^2}{B^2}+1\right],\label{Mdef}
 \end{equation}
that is characteristic of spherically symmetric spacetimes. Notice that this function furnishes an expression for $W$ through the constraint 
(\ref{WMconstr}) that allows us to eliminate the evolution equation (\ref{dotW}) for $\dot W$ (and $W$ in (\ref{dotSig})), though, since $M$ is fully 
expressible through (\ref{Mdef}) in terms of the variables of (\ref{1plus3EFE}), we do not need to use this function explicitly to integrate this system 
(we just eliminate $M$ with (\ref{Mdef}))\footnote{
To work beyond spherical symmetry we can easily do away with the usage of the Misner--Sharp mass function and work 
with the electric and/or magnetic Weyl tensor.}.
Besides these constraints, we also need to supplement the system with the conservation equation for each fluid:
\begin{subequations}
\begin{eqnarray}
\dot \rho_0 &=& -3 \left(\rho_0^\ast +p_0^\ast \right)H,\qquad A=-\frac{p_{0,r}^\ast}{\rho_0^\ast+p_0^\ast}, \label{eqn:Apr-rhoP}
\\
\dot \rho_{(i)} &=&- 3\left(\rho_{(i)}+p_{(i)}\right)H-2 \frac{A Q_{(i)}}{B^2}-6 \Pi_{(i)} \Sigma
-2\frac{Q_{(i)} \chi}{Y B^2}
\nonumber
\\
&{}&\qquad\qquad\qquad\qquad\qquad-\frac{Q_{(i),r} }{B^2}+\frac{Q_{(i)} B_{,r}}{B^3},\label{eqn:RhoiGen}
\\
\dot Q_{(i)}&=&-3 H Q_{(i)}-p_{(i),r}-A \rho_{(i)}-A p_{(i)}+2 \Pi_{(i)} A+2\Pi_{(i),r}
\nonumber
\\
&{}&\qquad\qquad\qquad\qquad\qquad\qquad+ \frac{6  \Pi_{(i)} \chi}{Y}.\label{eqn:QiGen}
\end{eqnarray}
\end{subequations}
Finally, the radial component velocity $V_{(i)}$ of the i--th fluid in (\ref{qu_def}) can be determined algebraically from (\ref{RhoPerFluid_i}) and (\ref{qPerFluid_i}).


\section{Void evolution from a mixture of two decoupled dusts}\label{Sec:VoidFormation}

The main characteristic of cosmic voids 
is the underdensity profile that depends on the (roughly) radial distance on 
Mpc scales. 
However, the usual single fluid (dust) approach generally focuses on the void 
dimensions (size and the depth of the density contrast) and the value of the local expansion \cite{BCK2011}, while the relative velocity between the dynamically significant 
species is usually ignored.
In this section we 
show 
that the multiple 
components scenario brings important modifications to the evolution of cosmic voids.

\subsection{A numerical example of the two--component mixture}
To stress the above it is illustrative to look at the case of a mixture of two dust fluids identified as CDM and 
baryonic matter, including a cosmological constant characterized 
by the present--day parameters from Planck 2015~\cite{ade2016planck}, in order to accommodate a $\Lambda$CDM asymptotic background model.
We consider the fundamental observers comoving with dark matter $u^\mu_{\DM}=\delta^{\mu}_{~t}$; 
consequently, the baryonic matter will have 4--velocity,
\begin{equation}\label{4v2ndDust:2dustproblem}
u^\mu_{\B}=\gamma \left(u_{\DM}^\mu+v^\mu\right), \;\;  \hbox{with} \;\; \gamma =\left(1-v^2\right)^{-\frac{1}{2}}, \;\;  \hbox{and} 
\;\; v_\mu=V \delta^{r}_{~\mu},
\end{equation}
and the energy--momentum tensor, eq. (\ref{Tequal_Sum_Ti}), will be the sum of the CDM and baryonic components:
\begin{subequations}\label{DMyBEoS}
\begin{equation}
T^{\mu\nu}= T^{\mu\nu}_{\DM}+T^{\mu\nu}_{\B}=\rho\, u^\mu u^\nu + p h^{\mu\nu} + 2 q^{(\mu} u^{\nu)} + \pi^{\mu\nu},
\end{equation}
with
\begin{equation}
 T^{\mu\nu}_{\DM}= \rho_{\DM} u^\mu u^\nu \;\;  \hbox{and} \;\;  T^{\mu\nu}_{\B}=\rho_{\B} u^\mu u^\nu + p_{\B} h^{\mu\nu} + 2 q_{\B}^{(\mu} u^{\nu)} + \pi_{\B}^{\mu\nu},
\end{equation}
%
such that one can identify the following quantities (as defined in~(\ref{subeqnTotalTherm}) and~(\ref{qu_def})), 
\begin{eqnarray}
\rho=\rho_{\DM} + \rho_{\B}, \qquad   p\equiv p_{\B}=\frac{1}{3} \gamma^2  v^2  \rho^\ast_{\B},  
\\
Q\equiv Q_{\B} = \gamma^2 \rho^\ast_{\B} V, \label{eqnQb}
\qquad 
\Pi\equiv \Pi_{\B}=\frac{1}{3}\gamma^2 \rho^\ast_{\B} v^2,
\\
\rho_{\DM}\equiv \rho^\ast_{\DM}, \qquad \rho_{\B}=\gamma^2 \rho^\ast_{\B}. \label{eqnRhob}
\end{eqnarray}
\end{subequations}
Note that 
the pressure, heat flow and anisotropic stress terms are zero when $V = 0$ (CDM and baryons with common 4--velocity: LTB limit). Consequently, the evolution will be governed by the system of equations 
that results from substituting $0\rightarrow\hbox{DM}$ and $i\rightarrow\hbox{B}$  in eq.~(\ref{1plus3EFE})
and considering the energy--momentum tensor variables (\ref{DMyBEoS}). The resulting dimensionless  system of equations  
is  presented explicitly in \ref{App:Dimensionless_system}.

\begin{figure*} 
\centering 
\includegraphics[width=.33\textwidth, clip]{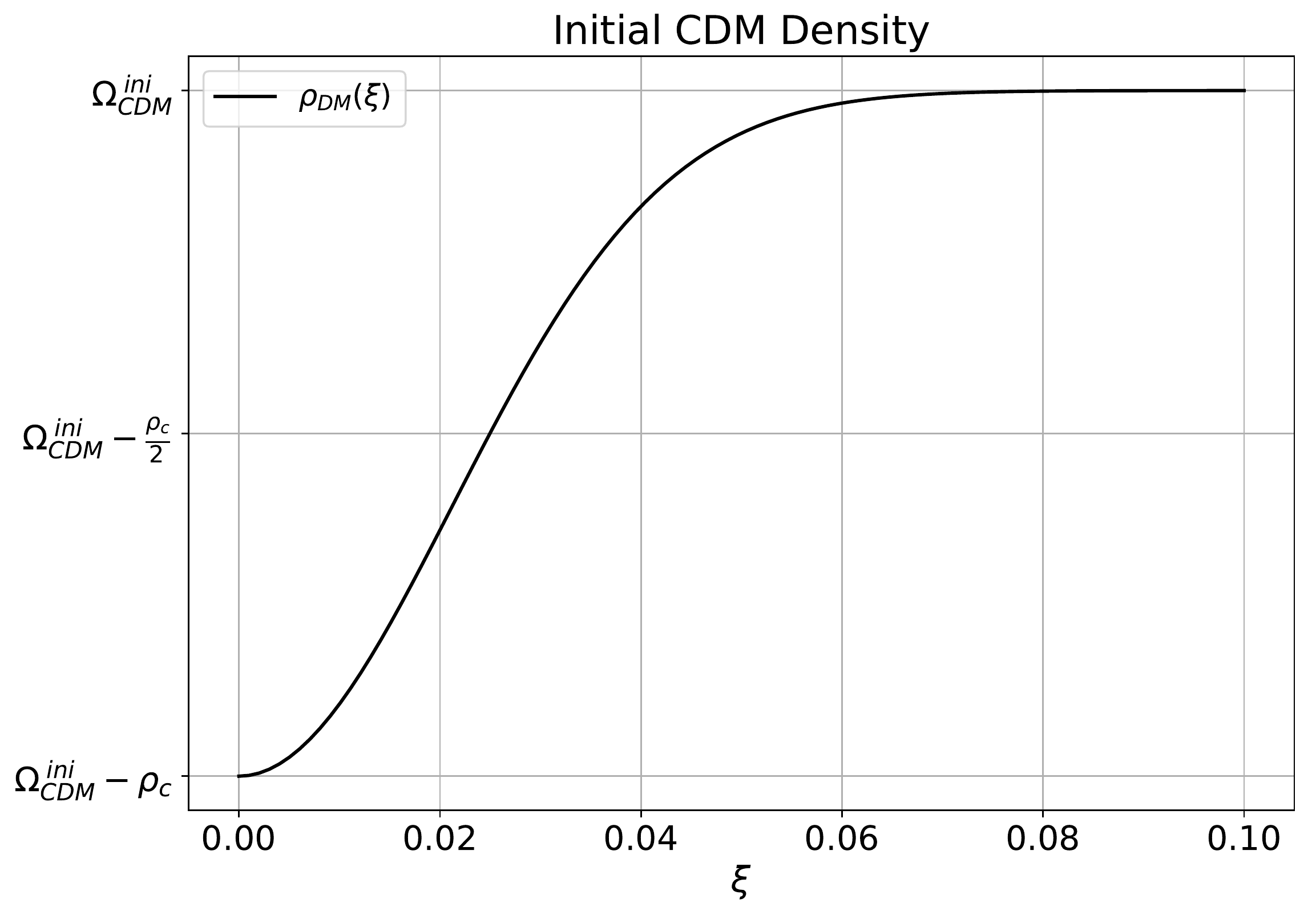}
\hfill
\includegraphics[width=.32\textwidth,clip]{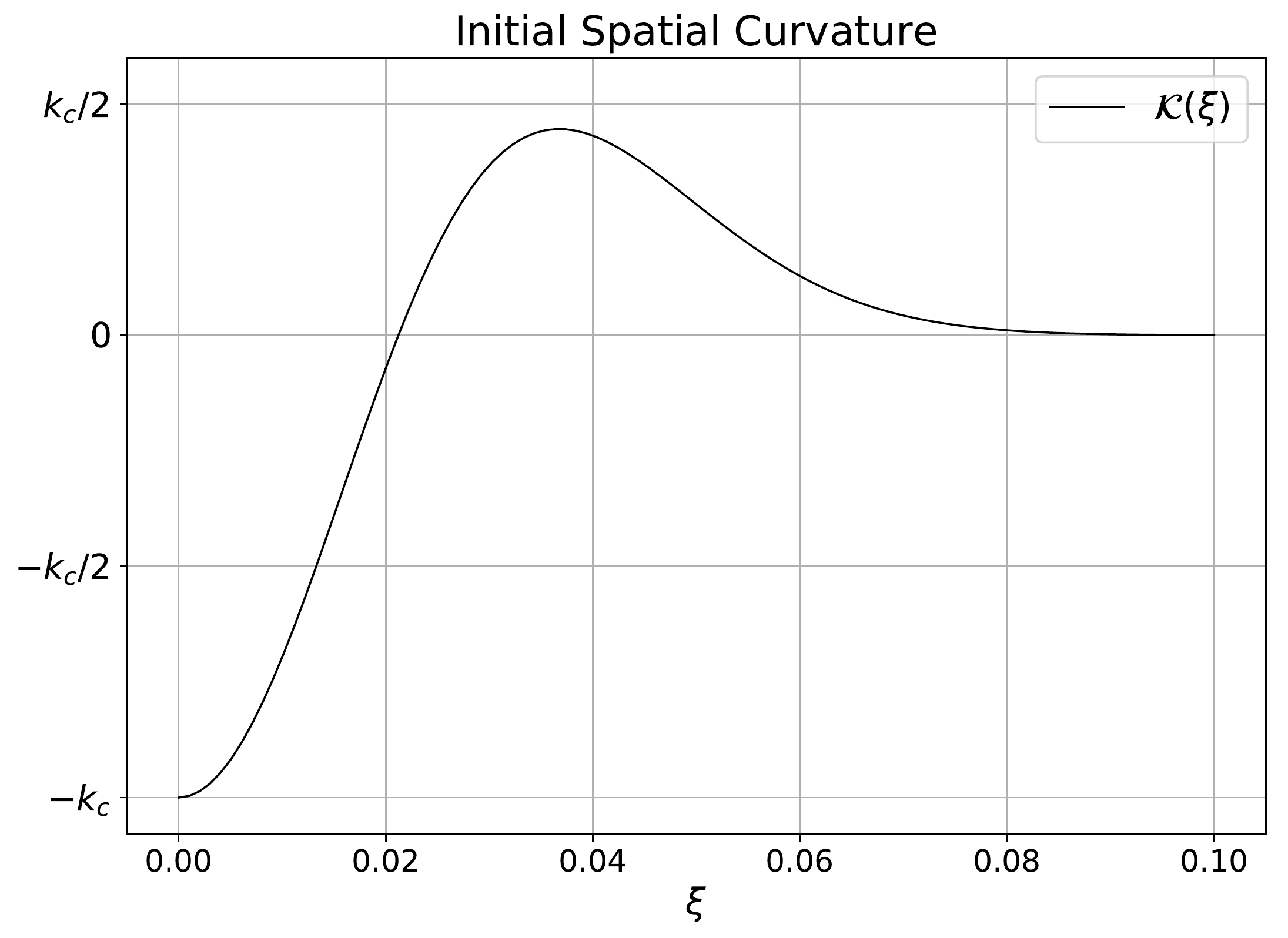}
\hfill
\includegraphics[width=.32\textwidth,clip]{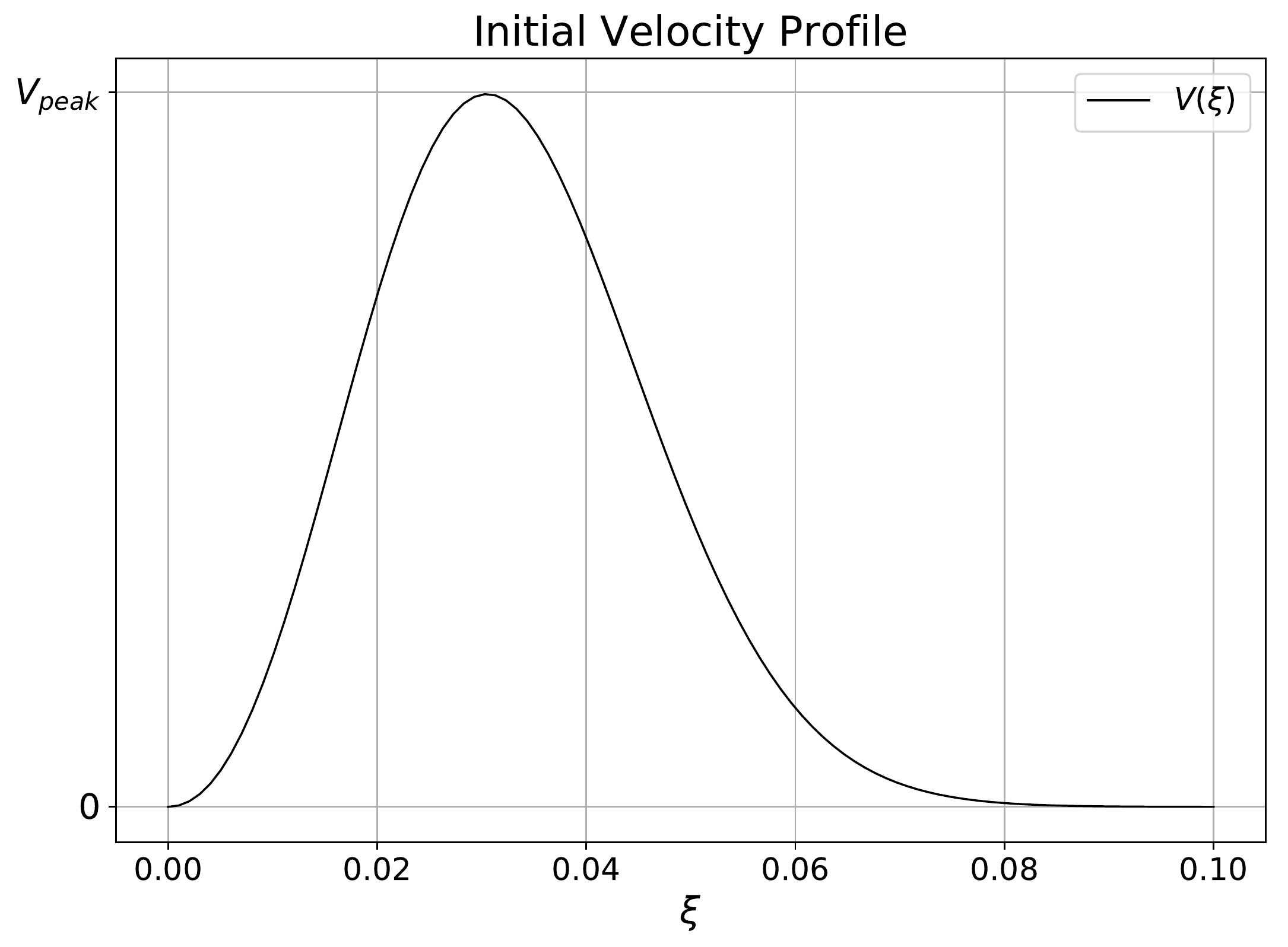}
\caption{\label{fig:InitProf} Initial CDM density, spatial curvature and baryon relative velocity profiles as functions of $\xi=r/l_\ast$. The initial functions are taken as Gaussian perturbations 
to the background functions, given by eq.~(\ref{IC:gf}) with $\mu_c\sim0.01$, $k_c\sim0.05$, $\sigma_K=\sigma_\mu=0.03$,
$\sigma_v=0.025$, while $V_{\peak}$ is varied over values between $-10^{-3}$ and $\sim10^{-2}$. The initial baryon matter density is homogeneous, as seen from the baryons intrinsic frame.
}
\end{figure*}

We examine the numerical solutions of this two--dust system in a grid simulating a cosmic void of present--day radius $\sim 60$ Mpc. Starting from linear initial conditions at 
 $z=23$ we follow its evolution until $z=0$ (see below for the justification of this choice of initial redshift). 
The initial CDM density, spatial curvature, and the relative velocity profiles are taken as Gaussian functions of linear amplitude with respect to the background parameters. In all our simulations the baryonic density  is initially homogeneous and equal to its value in the background (as seen in its intrinsic frame),
\begin{subequations}\label{IC:gf}
\begin{eqnarray}
\left[\frac{8 \pi}{3} \frac{\rho_{\DM}}{H^2}\right]_{\ini}&=&\Omega_{\DM}^{\ini} -\mu_c \exp\left({-\frac{\xi}{\sigma_\mu}}\right)^2, 
\\
\left[ K \right]_{\ini}&=& -\textit{k}_c \xi^2 \exp\left({-\frac{\xi}{\sigma_K}}\right)^2, \label{IC:K}
\\
\left[  V \right]_{\ini} &=& V_{{{c}}} \xi^2 \exp\left({-\frac{\xi}{\sigma_v}}\right)^2.
\end{eqnarray}
\end{subequations}
In the expressions above $\mu_c\sim0.01$, $\textit{k}_c\sim0.05$, $\sigma_K=\sigma_\mu=0.03$,
$\sigma_v=0.025$, and $r=l_\ast \xi$. From this, the spatial curvature ($\mathcal{K}$) is derived from Eqs.~(\ref{KurvEqn}) and~(\ref{IC:K}). The characteristic length is $l_\ast \sim 60$ Mpc, while the characteristic speed constant $V_c$ (and the maximum of the velocity $V_{\peak}$) will be specified further below.
Figure \ref{fig:InitProf} shows the 
typical initial profiles used for the numerical analysis.


\subsection{Evolution of density profiles}
To look at the effect of the relative velocity on voids and wall formation we develop a code capable of handling test cases with given initial densities for each species, a given curvature profile, and a series of profiles for the relative velocity ($V_{\peak}$). 

\begin{figure}
\centering 
\includegraphics[width=.40\textwidth,clip]{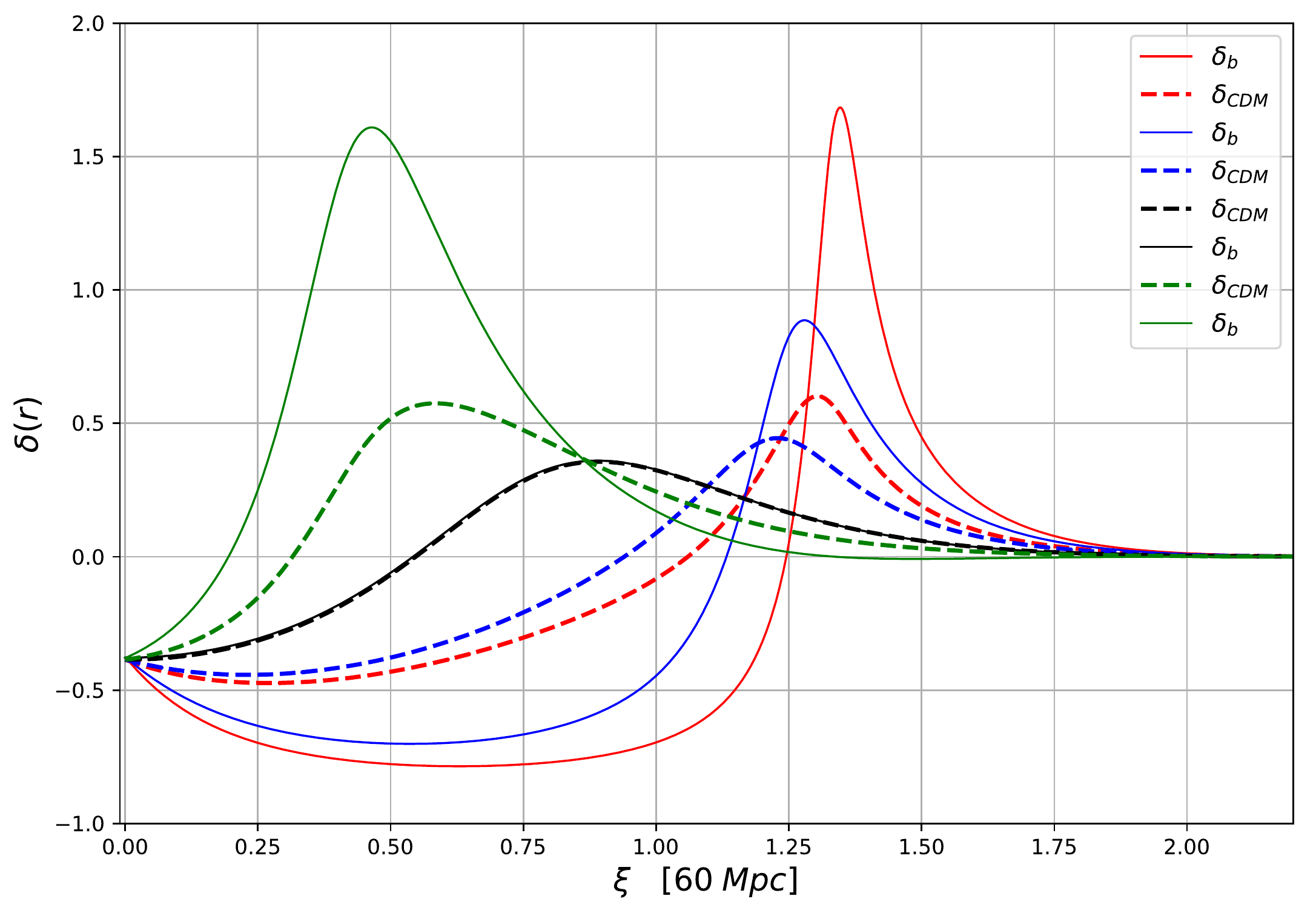}
\caption{\label{fig:DC} 
Influence of the initial baryon peculiar velocity on voids and wall formation. Baryon and CDM density 
contrasts ($\delta_{(i)}=\rho_{(i)}/\bar{\rho}_{(i)}-1$, where $\bar{\rho}_{(i)}$ is the value of $\rho_{(i)}$ in the background and $i=DM,B$) are depicted by dashed and solid lines, respectively, for different initial peculiar velocity profiles: all of them 
Gaussian functions with different amplitude ($V_{\peak}$ in Fig.~\ref{fig:InitProf}) set initially at $z=23$.  The red lines stand for $V_{\peak}\sim 7\times 10^{-3}$, the blue lines for $V_{\peak} \sim 5 \times 10^{-3}$ and the green ones for 
$V_{\peak}\sim -2.6 \times 10^{-3}$.
As a reference, we have provided the case without a relative velocity (LTB model), denoted by a black line. 
Note that the solution displayed by the solid blue curve includes a baryonic matter shell of width  $\sim 10$ Mpc 
and density contrast of the order the unity and peculiar velocity of $\sim 500$ km/sec with respect to the CDM comoving frame. 
This configuration is roughly comparable with the dynamics of our local group,  which has similar size and density contrast
and a dipole velocity of $\sim 600$ km/s associated with its local motion with respect to the CMB frame  
 \cite{DipoleVelCMB}.
} 
\end{figure}

\begin{figure*}
\centering 
\includegraphics[width=.4\textwidth, clip]{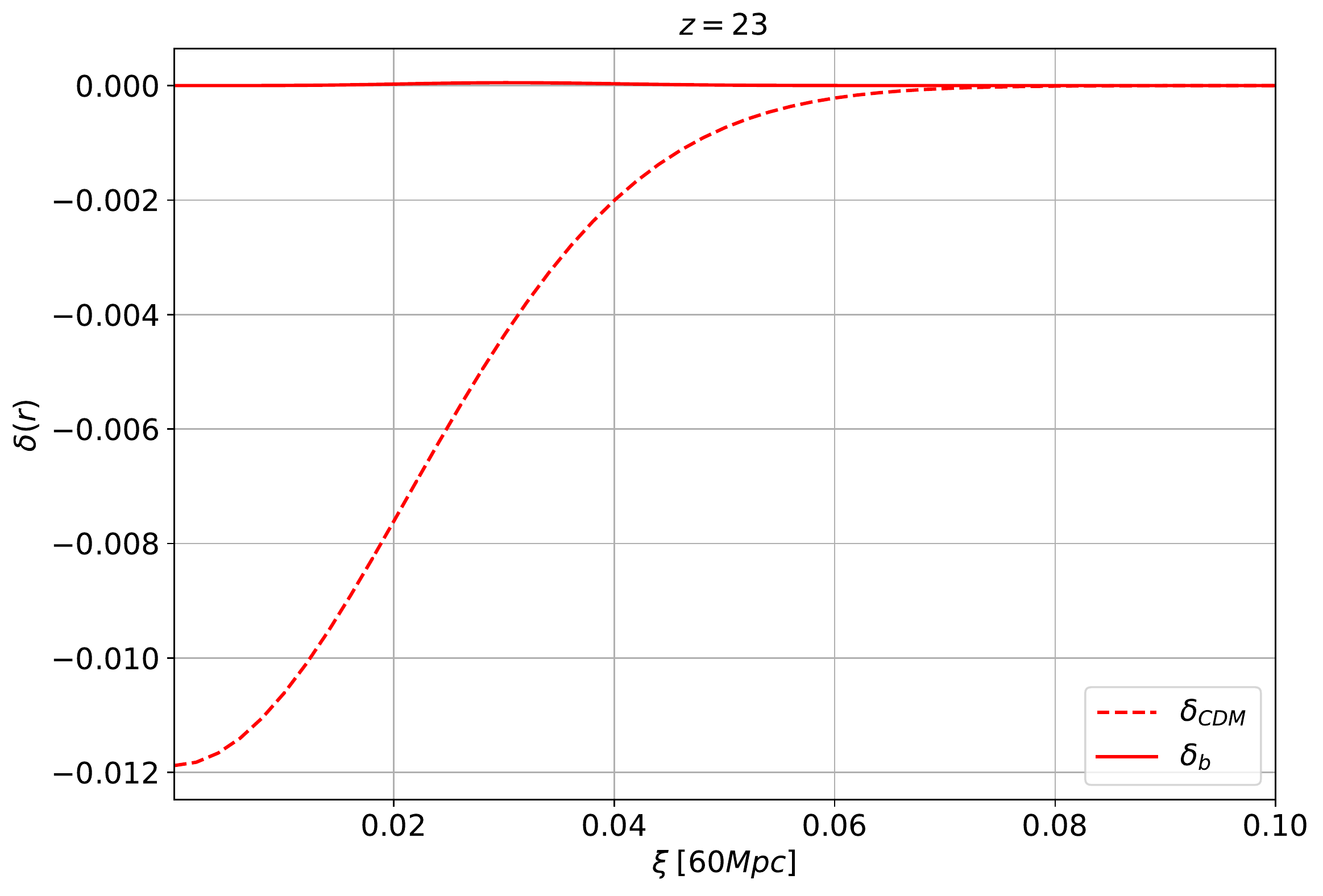}
\hfill
\includegraphics[width=.4\textwidth,clip]{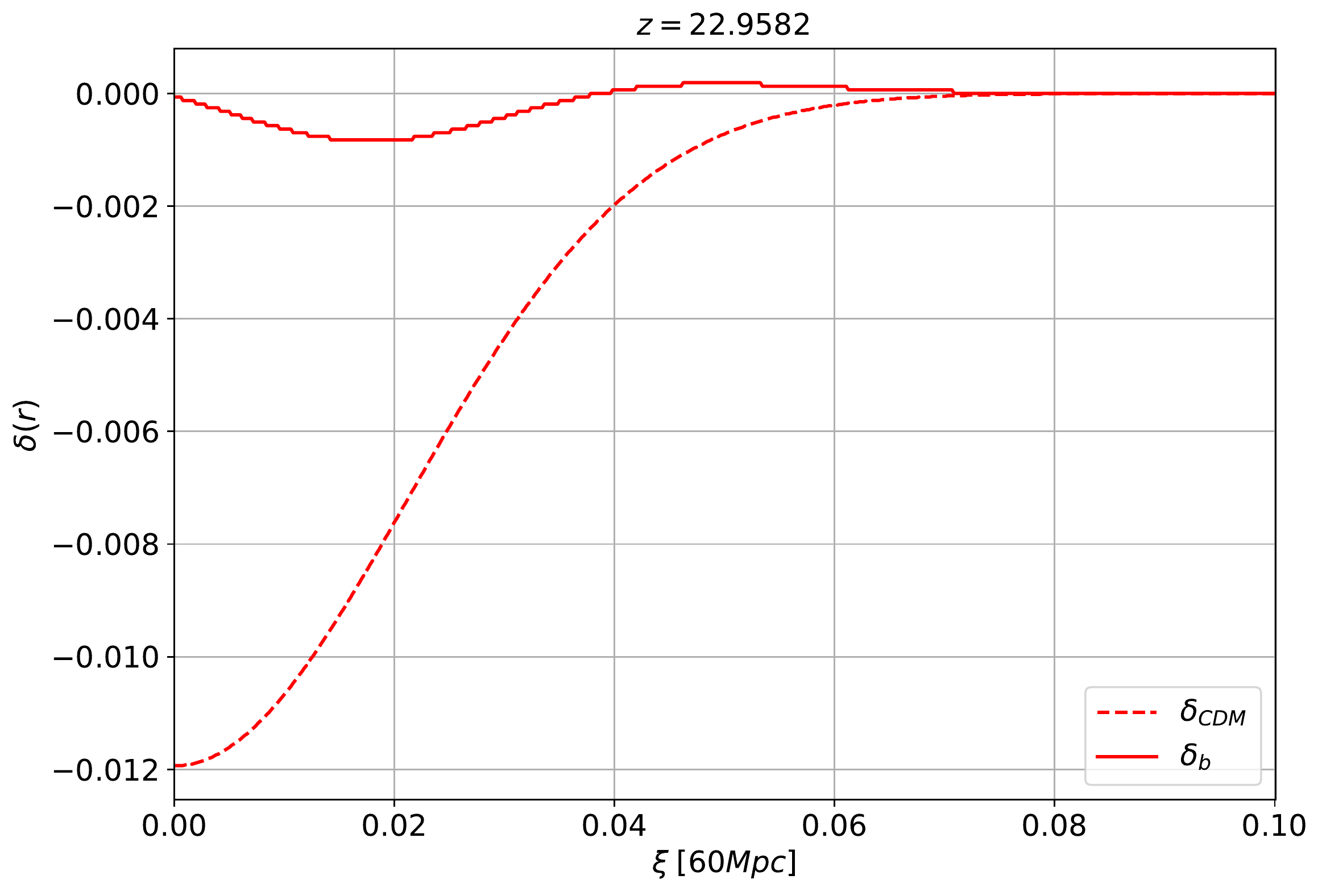}
\hfill
\includegraphics[width=.4\textwidth,clip]{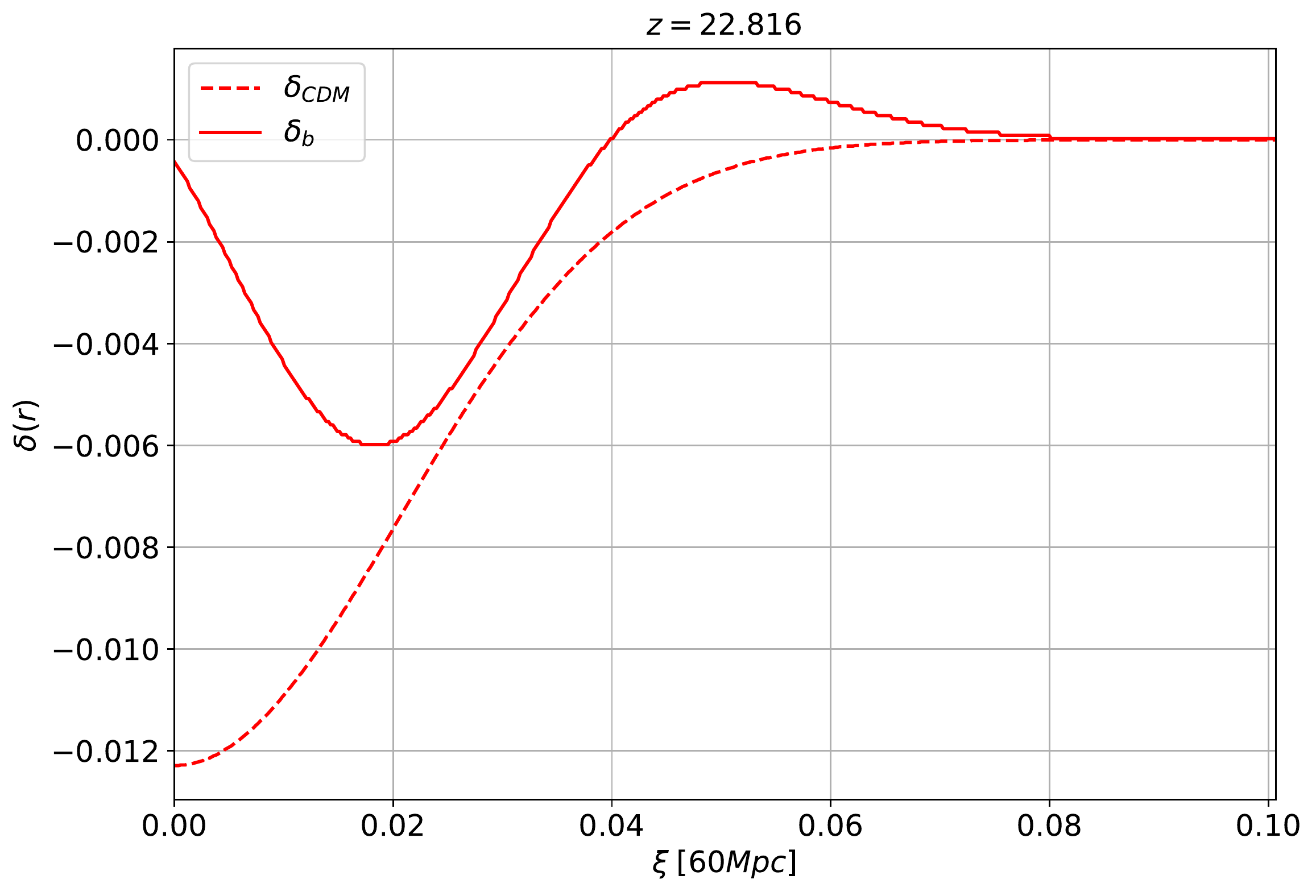}
\hfill
\includegraphics[width=.4\textwidth,clip]{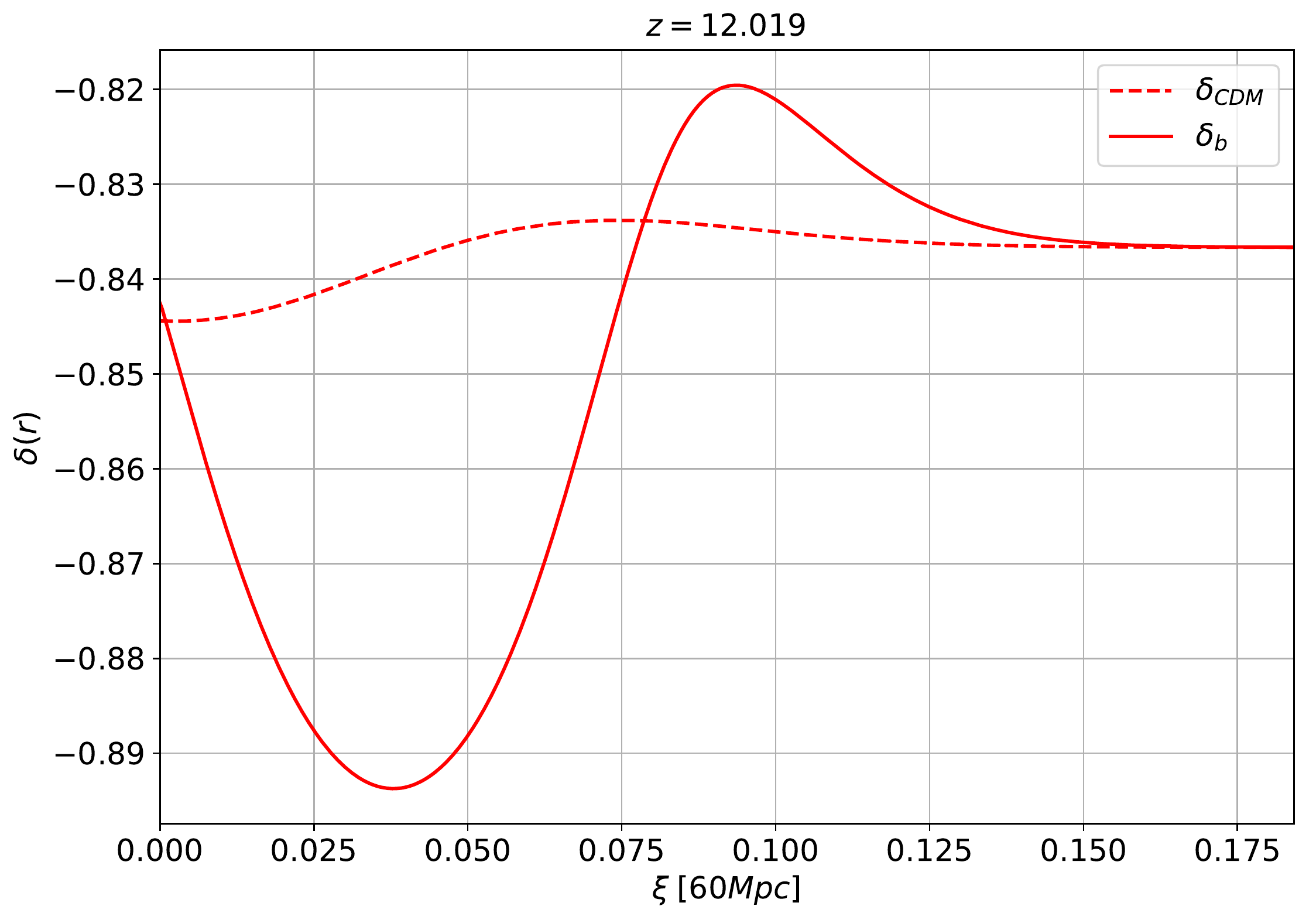}
\hfill
\includegraphics[width=.4\textwidth,clip]{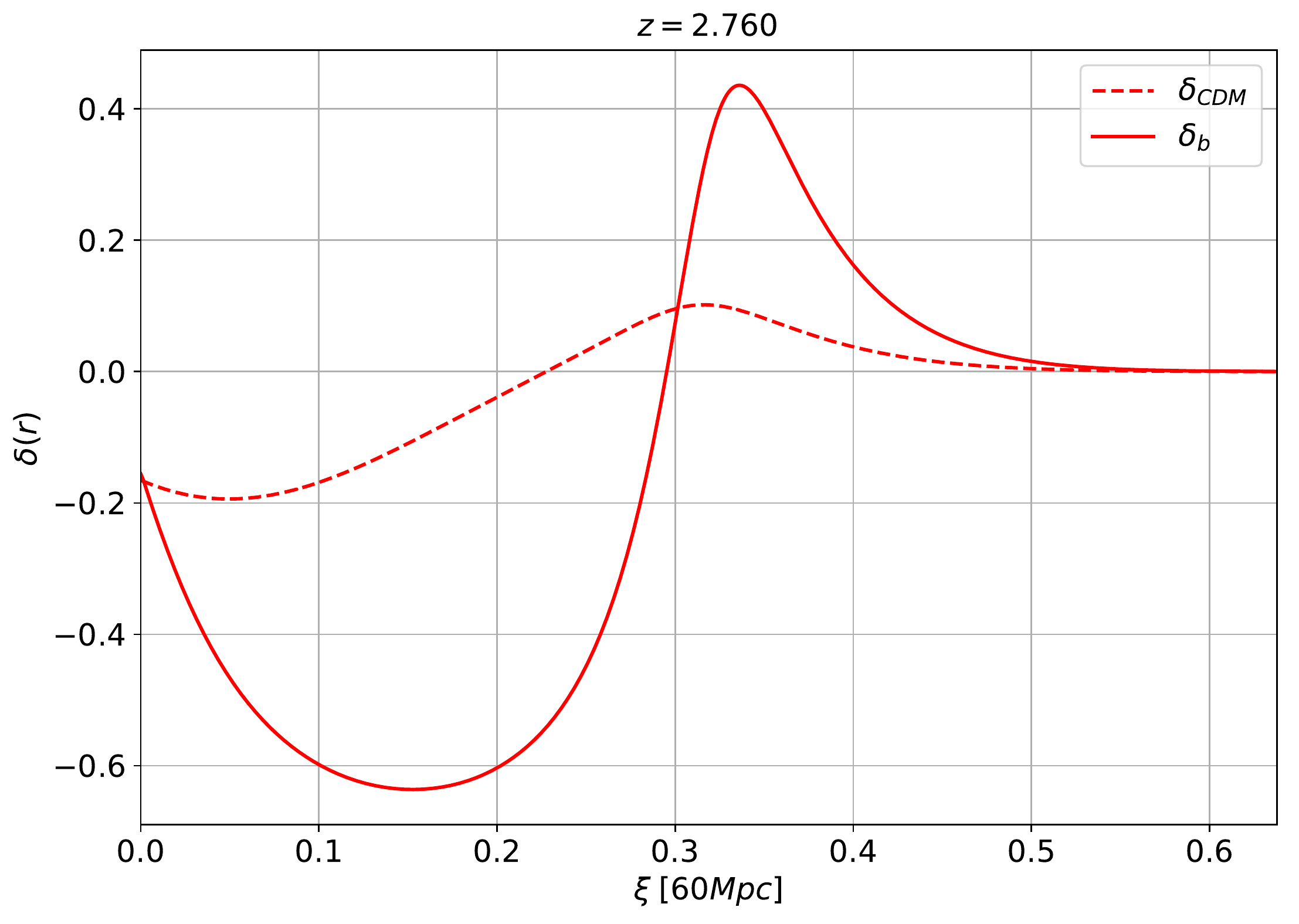}
\hfill
\includegraphics[width=.4\textwidth,clip]{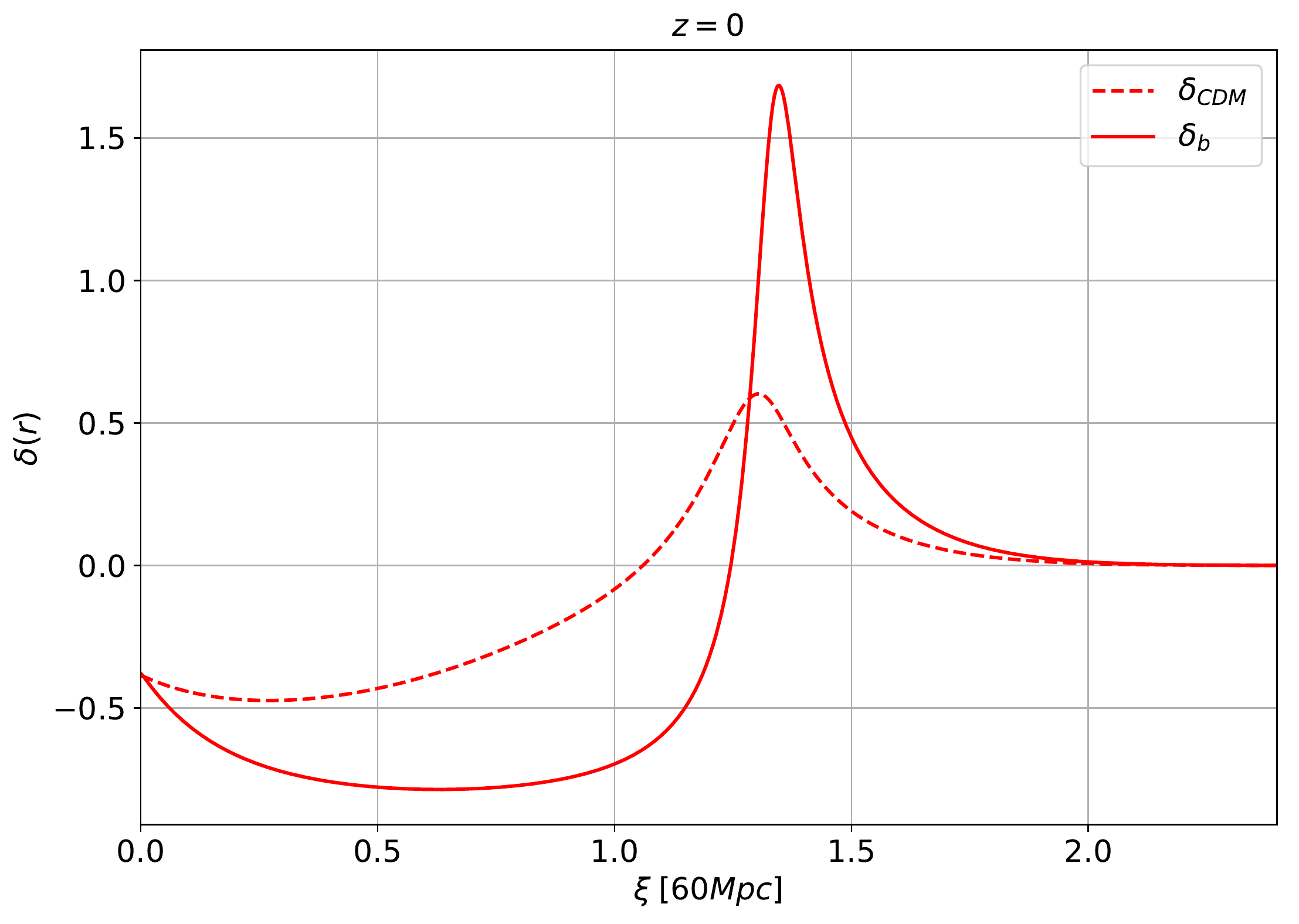}
\caption{\label{fig:SnapShots} 
The figure shows snapshots of the density contrast of each matter component at different redshifts for the solution depicted with red lines in Fig. \ref{fig:DC}. 
}
\end{figure*}

In  Fig. \ref{fig:DC} we display the baryon and CDM density contrasts at $z=0$ for different initial velocity profiles. As a reference, we have 
included the case in which both baryons and CDM are comoving (LTB solution).
We find that even non--relativistic relative velocity values exert non--trivial effects on present--day configurations 
as density contrasts become non--linear. 
On the other hand, 
the void size depends on the sign of $V$, so that smaller voids result from initially negative values for the relative velocity.
We also illustrate the evolution of the density contrast profiles for the specific case of $V_{\peak}\sim 7\times 10^{-3}$ 
(corresponding to the red curves in Fig. \ref{fig:DC}), snapshots for different values of $z$ are displayed in Fig. \ref{fig:SnapShots}.

Notice that as the evolution proceeds the density contrast at the surrounding wall increases, reaching probably a 
shell--crossing singularity. We interpret this as the onset of an intricate virialization process, a stage of structure formation  that marks the limit of validity of the dust model, and that lies beyond the scope of this work (discussed 
elsewhere in the literature~\cite{2008gady.book.B,PhysRevD.97.104029}). Since our purpose is to look at the simultaneous evolution of the 
matter--energy components (CDM and baryons) within the void before the onset of virialization,  we have 
%
chosen $z=23$ as the initial time slice, simply because it is easier to set the initial conditions at this time than at, say, the linear regime of the 
last scattering time $z\simeq 1100$, well before gravitational clustering becomes dynamically significant.  
However, these initial conditions are idealized but not fine--tuned or unrealistic, they simply correspond to a spherically symmetric realization of the generic spectrum of random CDM and baryon perturbations, characteristic of the linear regime at the last scattering surface $z \simeq 1100$, which evolve to produce a void of the desired size.


\subsection{Local expansion of the components}
Let us now focus on the effects of a relative velocity in the measure of kinematic quantities. In the case of a single fluid, the comoving observers define a natural threading of the spacetime by the future--directed unit timelike
vector field $u^\mu$. 
In our case, as we stressed before, the choice of the fundamental observers is not unique, 
and observers comoving with each fluid will measure the kinematic quantities with different magnitudes. In fact, 
due to a change of frame, the local expansion of CDM ($H_{\DM}\equiv H$ in this frame) will depart from the expansion of the baryonic matter ${H_{\B}}$, which is given by
\begin{equation}
3 {H_{\B}}=\Theta_{\B}={h_{\B}}_{\mu}^{\,\nu} \nabla_{\nu} u^\mu_{\B},\label{Theta_b:eq1}
\end{equation}
where ${h_{\B}}^{\mu\nu}=u^\mu_{\B} u^\nu_{\B} + g^{\mu \nu} $ is the projection tensor and $u^\mu_{\B}$ is the 4--velocity of the baryonic matter given in eq.~(\ref{4v2ndDust:2dustproblem}). We find by computing (\ref{Theta_b:eq1}) a relation between both estimations of the expansion,
\begin{eqnarray}
3{H_{\B}}&=&\left[\left(\frac{2 Y_{,r}}{Y}-\frac{B_{,r}}{B}\right) \frac{V}{B^2}+\frac{V_{,r}}{B^2}+3 H_{\DM}\right] \gamma+\frac{V}{B^2}\gamma_{,r} +\dot\gamma,\nonumber
\\
&\simeq & \left(\frac{2 \chi}{Y}-\frac{B_{,r}}{B}\right) \frac{V}{B^2}+\frac{V_{,r}}{B^2}+3 H_{\DM}+V \dot V,\label{Theta_b:eq3}
\end{eqnarray}
where in order to derive eq.~(\ref{Theta_b:eq3}) we have used the fact that $V\ll 1$ (but its derivatives need not be small) and  $\chi\equiv Y_{,r}$. Hence, the difference in the local expansion due to a change of frame can be expressed as follows, 
\begin{equation}
3 \left({H_{\rm B}}-H_{\DM}\right) \simeq 
 \left(\frac{2 \chi}{Y}-\frac{B_{,r}}{B}\right) \frac{V}{B^2}+\frac{V_{,r}}{B^2}+V \dot V.
 \label{diff:Th-Th_b}
\end{equation}
\begin{figure*}
\centering 
\includegraphics[width=.40\textwidth,clip]{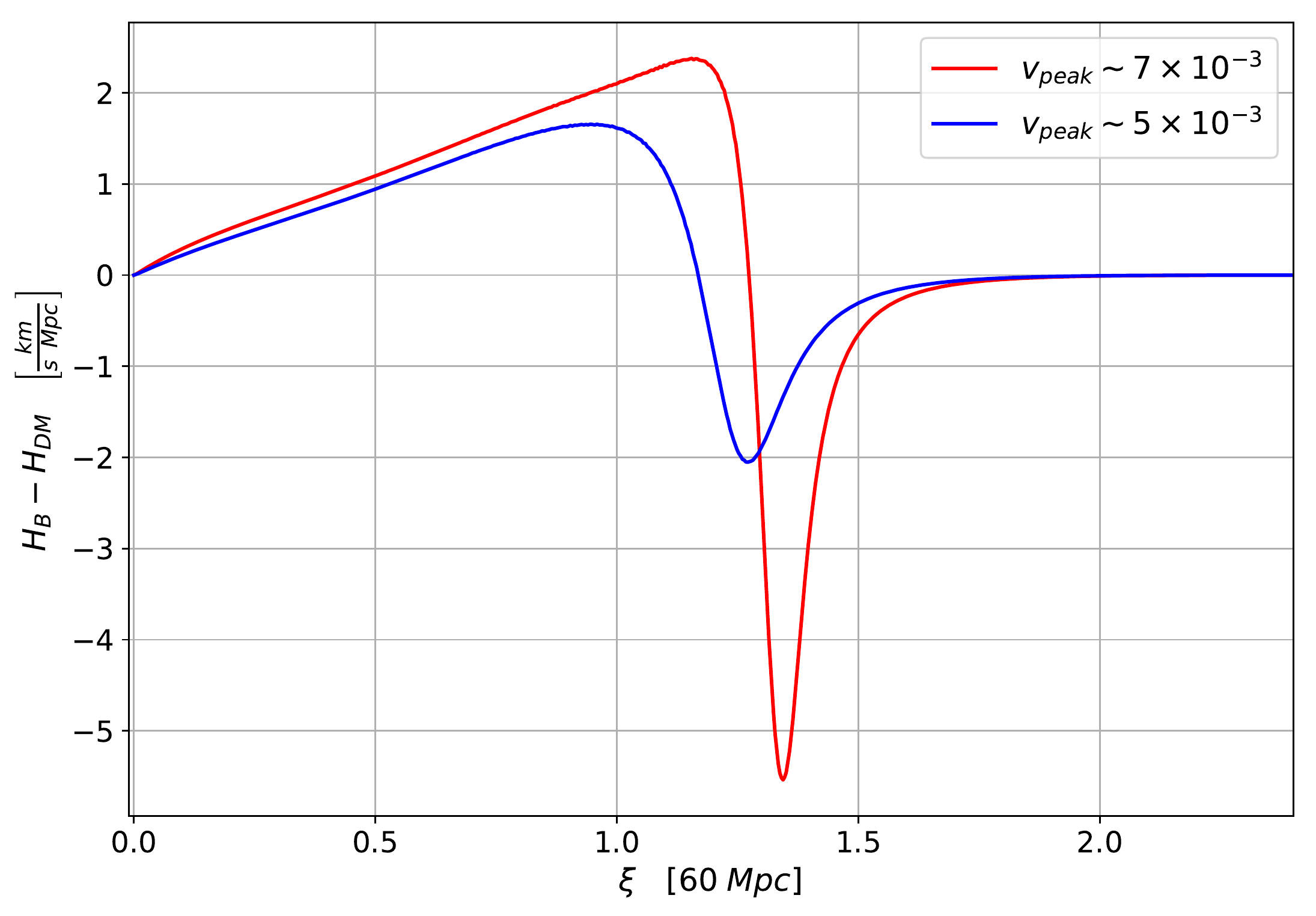}
\hfill
\includegraphics[width=.40\textwidth]{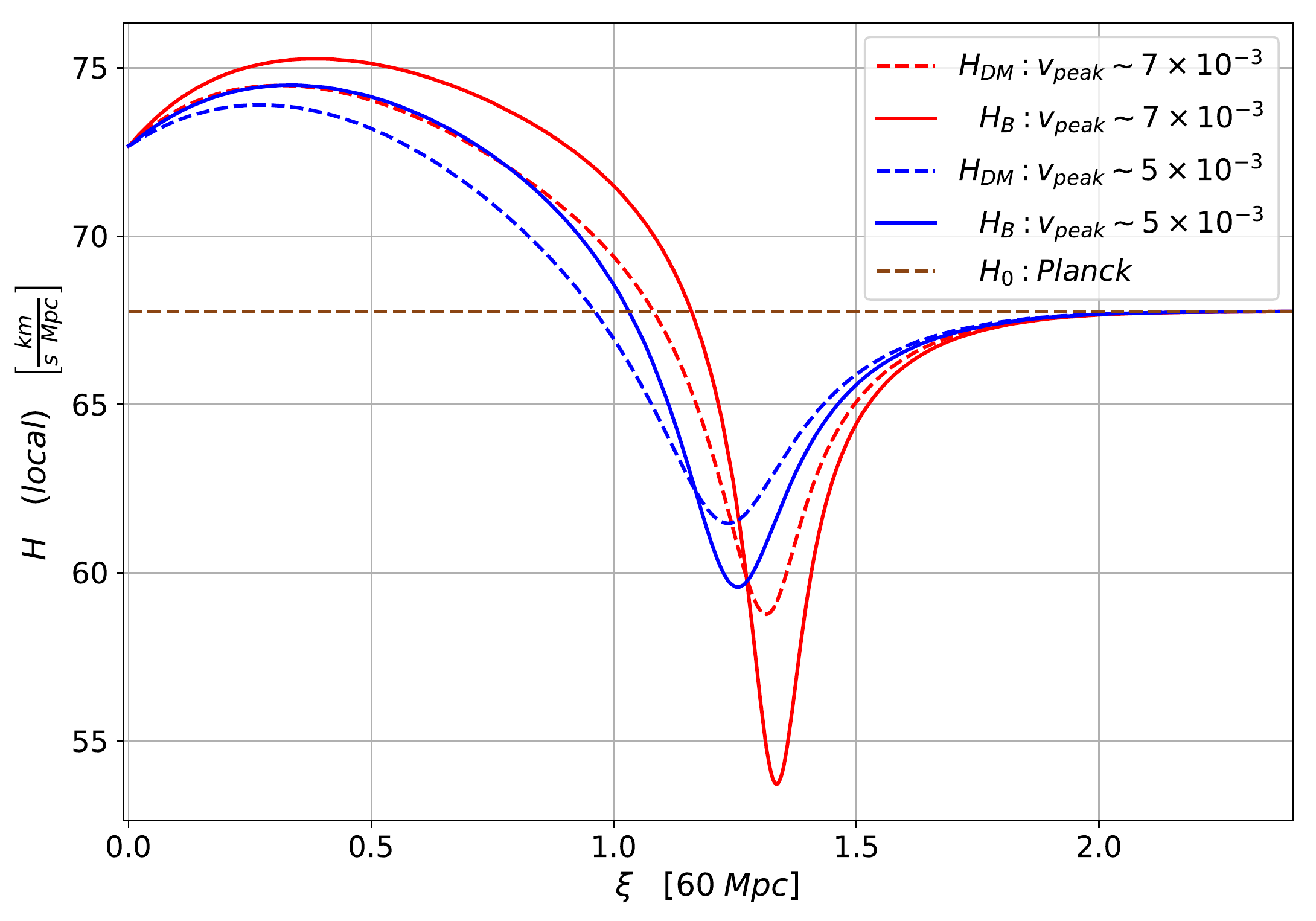}
\caption{\label{fig:dH} Differences in the local expansion due to a change of frame. The left panel shows the difference between the two expansions at $z=0$ (Eq.~\eqref{diff:Th-Th_b}) for two of the solutions depicted in Fig.~\ref{fig:DC}, corresponding to $V_{\peak}$ equal to $7\times10^{-3}$ (red lines) and $5 \times 10^{-3}$ (blue lines). The right panel shows the local expansion of each component as computed from the solutions of the system~\eqref{2DustSysEqns} and Eq.~\eqref{diff:Th-Th_b}.
}
\end{figure*}
%
%
Figure \ref{fig:dH} shows the difference between the two expansions $H_{\rm B}$ and $H_{\DM}$ at $z=0$ for the solutions whose density contrast
are depicted with red and blue lines in Fig.~\ref{fig:DC}, corresponding to $V_{\peak}$ equal to $7\times 10^{-3}$ (red lines) and $5 \times 10^{-3}$ (blue lines).
Note that this difference can be of the order of $ \mathrm{km/ (s \,  Mpc)}$, around the maximum of the baryonic matter density (even larger differences are expected at times close to virialization). 
This estimation is roughly that of the discrepancy between the values of $H_0$ reported by CMB and SNe observations~\cite{RiessH02016,ade2016planck,freedman2017cos}, thus suggesting that considering a relative velocity between baryons and CDM may provide interesting clues to understand this issue (though this task lies beyond the scope of the present work). 
%

\subsection{The baryon--CDM relative velocity}

Since for the baryon--CDM mixture the radial component of the relative velocity, as defined in Eq.~\eqref{4v2ndDust:2dustproblem}, can be 
determined from the algebraic relation
\begin{equation}
V=Q_{\B}/\rho_{\B}, 
\end{equation}
with $Q_{\B}$ and $\rho_{\B}$ given by Eq.~\eqref{eqnQb}--\eqref{eqnRhob}, its evolution equation is 
\begin{eqnarray}
\dot V &=& \left( H-2 \Sigma \right) \frac{V^3}{B^2}+\frac{B_r}{B^3} V^2-\frac{V_{,r}}{B^2}V \nonumber
\\
&=&\left( \frac{H-2 \Sigma}{B^2} \right) V^3-\left(\frac{V^2}{2 B^2}\right)_{\!\!\!,r}.
\end{eqnarray}
where we used~\eqref{eqn:RhoiGen},~\eqref{eqn:QiGen} and $A\equiv0$. 
In order to relate this equation to the well--studied perturbative case, we drop the 
term of order of $V^3$ to obtain,
\begin{equation}
\dot V \approx -\left(\frac{V^2}{2 B^2}\right)_{\!\!\!,r},
\end{equation}
which shows the connection between the time evolution of the relative velocity and the radial 
 gradients of the velocity and the metric function $B$ that generalizes the background scale factor. 
In a quasi--homogeneous perturbative regime $B\sim a(t,r)\,r$ and thus $B_{,r}>0$ should hold, while $V_{,r}>0$ 
should also hold because relative velocities increase from the center onwards as $r$ increases. Therefore the derivative of 
$(V^2)/(2B^2)$ should be positive and thus the right--hand side of the equation above negative. As a consequence, $\dot V<0$ 
holds and relative velocities dilute asymptotically during cosmic expansion. However, this is not applicable to a non--perturbative 
regime where large gradients of the involved variables may occur and/or change signs, so that the relative velocity can be amplified by a local inhomogeneity.

We can obtain further information on the evolution of $V$ by looking at a definition of peculiar velocity often used in a perturbative approach: 
the difference between the local Hubble flow relative to the Hubble flow of the background, which can 
be estimated as $v_{pec}=(H_{\tiny{\rm{local}}}-H_{\tiny{\rm{FLRW}}}) Y$ \cite{Bolejko2008}. Then, 
once again neglecting the highest power of $V$ in Eq.~\eqref{diff:Th-Th_b}, we get
\begin{equation}
\Delta v_{pec}=\left({H_{\B}}-H_{\DM}\right) Y \simeq  \left(\frac{2 \chi}{Y}-\frac{B_{,r}}{B}\right) \frac{Y}{B^2} V+\frac{Y}{B^2} V_{,r},
\end{equation}
which shows that such spatial velocity field is intrinsically related with the local homogeneities.  
At large scales (in a perturbative regime) this field evolves by approximately diluting  
as the inverse of the background scale factor, since in a regime approaching FLRW--like conditions our variables can be written 
as $B\sim a(r,t)$ and $Y\sim r a(r,t)$  (see e.g., \cite{Sussman:2014wua} for a formal equivalence of LTB models with Cosmological Perturbation Theory in the linear regime). In an inhomogeneity, however, where the spatial gradients are 
not restricted to small values, the relative velocity and peculiar velocity must be found by a non--trivial evolution equation.
 In particular, for the numerical solutions showed in this section we found that the relative velocity decreases, 
 but without following a trivial scaling law in the spatial region identified with the walls. 
 Note that in such regions the gradients can be large and the local expansion is slower than the background expansion, 
 in fact, in part it is locally collapsing.


\section{Discussion and final remarks}\label{Sec:Disc-Remarks}

We have considered the fully relativistic evolution of spherically symmetric cosmic voids made up of a mixture of two non--comoving 
dust components, 
identified as CDM and baryonic matter. Specifically, we looked at the effects of the baryon--CDM relative velocity on the void properties. We found that 
for baryons converging to the center of the void, as seen from the CDM frame, the final density profile shows an effective reduction on the size of the void (see Fig.~\ref{fig:DC}). 
On the other hand, if the baryon component is receding from the centre, the void presents a deeper (baryonic) underdensity, and the walls manifest a larger density contrast as illustrated in Figs.~\ref{fig:DC} and \ref{fig:SnapShots}. 

\begin{figure}
\centering 
\includegraphics[width=.45\textwidth,clip]{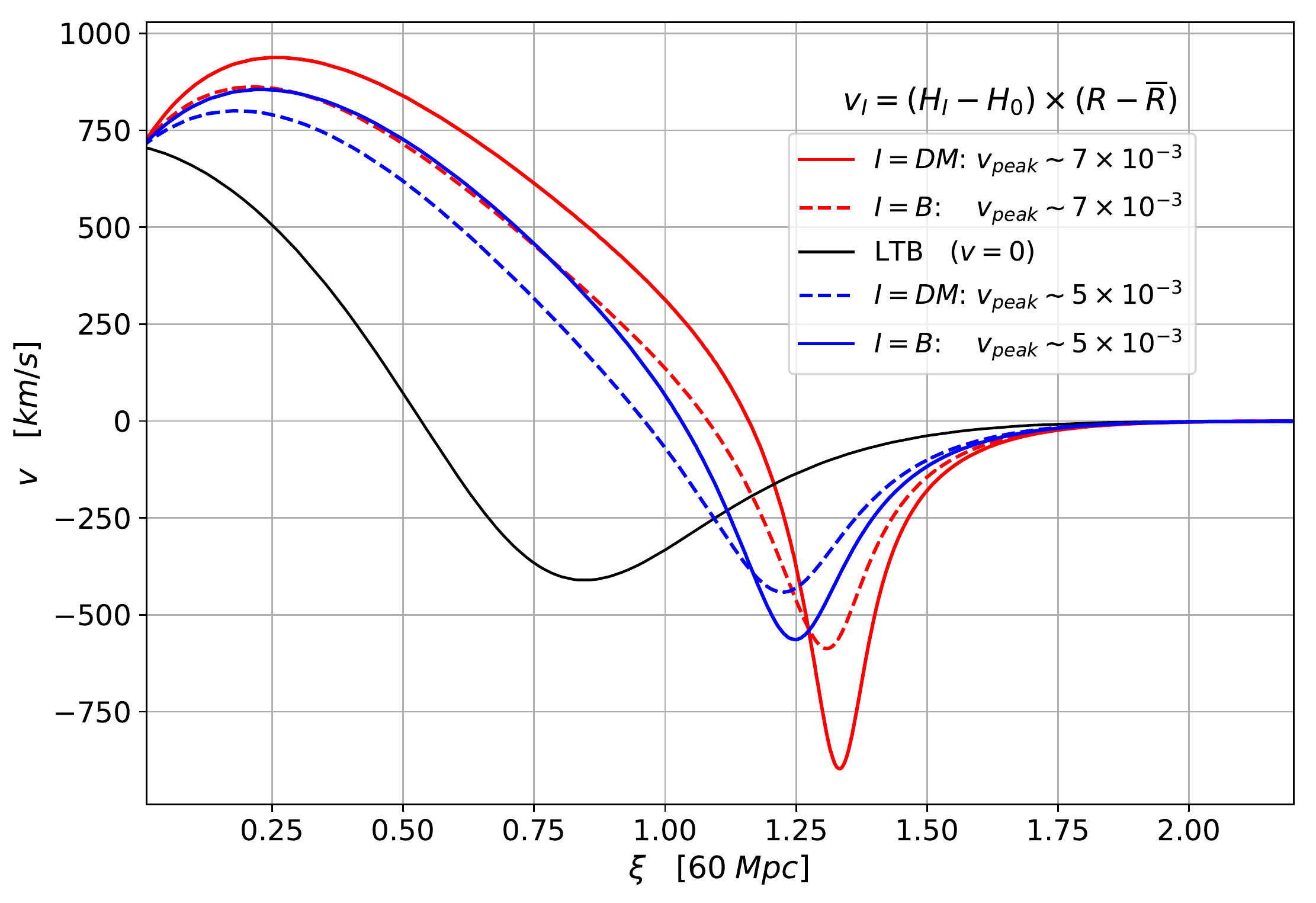}
\caption{\label{fig:peculiar} 
Peculiar velocity of the void components with respect to the asymptotic background as defined in  \cite{Bolejko2008}. 
Note that the larger differences occur at the radii corresponding to the wall structure.
}
\end{figure}

The existence of a relative velocity between baryons and CDM leads to a difference in the expansion of each component. We find that small initial differences in velocities between two components (of order $7$--$5\times 10^{-3}$) yield important differences in local expansion of the order  of  $\, \mathrm{km/ (s \, Mpc)}$, similar to the gap between local and CMB measurements of the expansion parameter $H_0$ ~\cite{RiessH02016,ade2016planck,freedman2017cos,macpherson2018trouble}. Indeed, this last result may be part of the effects missing in the usual single frame analysis 
of peculiar velocities and local expansion (e.g. curvature effects \cite{bolejko2017relativistic,PhysRevD.97.103529}, among others).
Related to this, we find significant differences in the peculiar velocities of each component, defined as deviations from the 
asymptotic \textit{background} 
(common) expansion, reached at large radii. 
Such differences could be interpreted as the velocity bias field, here evolved to non--linear stages. Fig.~\ref{fig:peculiar} shows that such bias manifests most prominently at the peak of the density contrast (walls of the void). 

The spherical void model we work with is qualitatively analogous to earlier models~\cite{kim2018alternative,tsizh2016evolution,novosyadlyj2016evolution,Marra:2011zp}. As in these models, we obtain 
qualitatively analogous results that depict the expected streaming of baryons determined by the rapid void expansion, a characteristic also present in
Newtonian models. However, the dynamical equations employed in the past are based on a 
numerical scheme constructed from the Misner--Sharp mass that is completely tied to spherical symmetry. As a contrast, the system of evolution 
equations and constraints here considered is based on covariant fluid--flow scalars that can be computed for any spacetime, regardless of 
the symmetry considerations. 


Our approach to void dynamics could also represent an important improvement over the 
``silent models'' of~\cite{bolejko2017relativistic} that try to address this 
issue through ``emergent'' spatial curvature. Silent models (characterized by a 
non--rotating dust source with purely electric Weyl tensor) are theoretically 
handicapped by the conjecture stating that Einstein's equations may not be integrable in general under the 
``silence'' assumption~\cite{van1997integrability,sopuerta1997new}  (the models in~\cite{bolejko2017relativistic} also neglect matching 
conditions among the different silent cells). By assuming dust sources with different (non--comoving) 4--velocities, the resulting 
models are based on similar physical assumptions but are no longer silent because of the non--trivial energy and momentum 
flux among the dust sources.

In conclusion, the fully relativistic evolution of baryons and CDM along different 4--velocity frames can provide important clues 
in understanding the observational tension in the estimation of the value of $H_0$ from local observations and through interpretation of the 
Planck data. 
A concrete example is furnished by the study of the Hubble flow in the non--spherical models 
examined in~\cite{macpherson2018trouble}, which tries to understand 
this tension, but did not consider different 4--velocities for the baryon and CDM components. 
This work could be improved by allowing for a non--comoving baryon 4--velocity that would 
provide more degrees of freedom as we have done in this paper. Likewise the multiple fluid approach can provide
important corrections to the usual study of the process of formation and growing of  large--scale structure in the universe.
Finally, we emphasize the fact that the system of evolution equations and constraints used in our numerical modeling has been
constructed with covariant fluid flow variables, and thus it is readily applicable (under certain restrictions)   
to examine self--gravitating systems that are much less idealized that those under the assumption of spherical symmetry.

\begin{acknowledgements}
The authors acknowledge support from research grants
SEP--CONACYT 282569 and 239639. I.D.G. also acknowledges valuable discussions with S. Fromenteau.
\end{acknowledgements}


\appendix

\section{Einstein's field equations as a first order ``$1+3$'' system}\label{App1plus3}

Given a 4--velocity field, Einstein's field equations are equivalent to a set of evolution and constraint equations involving the kinematic parameters $\Theta,\,\dot u^\mu,\,\sigma_{\mu\nu},\,\omega_{\mu\nu}$ (expansion, 4--aceleration, shear and vorticity), the components of the energy momentum tensor $\rho,\,p,\Pi_{\mu\nu},\,q_\mu$ (energy density, isotropic and anisotropic pressure, energy flux) projected by the 4--velocity, as well as the electric and magnetic parts of the Weyl tensor $E_{\mu\nu},\,H_{\mu\nu}$. For spherical symmetry we have $\omega_{\mu\nu}=H_{\mu\nu}=0$, hence the 1+3 system becomes the evolution equations
\begin{eqnarray}
&{}& \dot \rho+3 \left(\rho+p\right) H+2\dot u^\mu q_\mu+\tilde\nabla_\mu u^\mu+\pi_{\mu\nu}\sigma^{\mu\nu}=0,\label{eqnAppdrho}
\\
&{}& 3\dot H+3 H^2+\frac{\kappa}{2}\left(\rho+3p\right)\sigma_{\mu\nu}\pi^{\mu\nu}-\tilde\nabla_\mu \dot u^\mu-\dot u_\mu\dot u^\mu=0,
\nonumber
\\
\\
&{}& \dot q_{\langle \mu \rangle}+4 H q_\mu +\left(\rho+p\right)\dot u_\mu+\tilde\nabla_\mu p+\tilde\nabla_\nu\pi^\nu_\mu+\pi_{\mu\nu}\dot u^\nu
\nonumber
\\
&{}& \qquad \qquad \qquad \qquad \qquad \qquad \qquad +\sigma_{\mu\nu}q^\nu=0,\\
&{}& \dot\sigma_{\langle \mu\nu\rangle}+2 H\sigma_{\mu\nu}+E_{\mu\nu}-\frac{\kappa}{2}\pi_{\mu\nu}-\tilde\nabla_{\langle \mu}\dot u_{\nu\rangle}
+\sigma_{\upsilon\langle \mu}\sigma_{\nu\rangle}^\upsilon
\nonumber
\\
&{}& \qquad \qquad \qquad \qquad \qquad \qquad \qquad -\dot u_{\langle \mu}\dot u_{\nu\rangle}=0,
\\
&{}& \dot E_{\langle \mu\nu \rangle}+3 H E_{\mu\nu}
+\kappa\dot u_{\langle \mu}q_{\nu\rangle}-3\sigma_{\upsilon\langle \mu}E_{\nu\rangle}^\upsilon
\nonumber
\\
&{}& \qquad+\frac{\kappa}{2}\left[\left(\rho+p\right)\sigma_{\mu\nu}+\dot\pi_{\langle \mu\nu\rangle}+\frac34 H\pi_{\mu\nu}+\tilde\nabla_{\langle \mu}q_{\nu\rangle}\right]=0,\nonumber\\
\end{eqnarray}
together with the constraints
\begin{eqnarray}
&{}& \tilde\nabla_\nu\sigma^\nu_\mu-2\tilde\nabla_\mu H+\kappa q_\mu=0,
\\
&{}&  \tilde\nabla_\nu E^\nu_\mu-\frac{\kappa}{3}\left(\tilde\nabla_\mu\rho-3 H q_\mu-\frac12\tilde\nabla_\nu\pi^\nu_\mu\right)
\nonumber
\\
&{}& \qquad \qquad \qquad \qquad \qquad \qquad \qquad -\frac{\kappa}{2}\sigma_{\mu\nu}q^\nu=0,
\\
&{}&  H^2-\frac{\kappa}{3}\rho-\frac16\sigma_{\mu\nu}\sigma^{\mu\nu}+\frac16\,{}^{(3)}{\cal R}=0, \label{eqnAppHamiltonianConst}
\end{eqnarray}
where  the ``dot'' and ``tilde'' respectively denote the convective (projected with $u^\mu$) derivative and spacelike gradients (projected orthogonal to $u^\mu$), see (\ref{eqn:convec-spacegrad}), while indices enclosed by angle brackets (${}_{\langle \mu\nu \rangle}$) denote the spacelike symmetric tracefree projection (see (\ref{TprojectionII})).

In order to apply the system (\ref{eqnAppdrho})--(\ref{eqnAppHamiltonianConst}) to the fluid mixture we need to substitute 
(\ref{RhoPerFluid_i})--(\ref{AniPPerFluid_i}) for the total forms of $\rho,\,p,\,\pi_{\mu\nu}$ and $q_\mu$, as well as the forms 
for the kinematic parameters and electric Weyl tensor in (\ref{acc_def}) and (\ref{eqn:sigma-EeigenV}).   

Notice that the system (\ref{eqnAppdrho})--(\ref{eqnAppHamiltonianConst}) is not only valid for spherically symmetric spacetimes, but for Petrov type D spacetime ($H_{\mu\nu}=0$) whose source is endowed with an irrotational fluid 4--velocity ($\omega_{\mu\nu}=0$).  The system can be readily extended to more general spacetimes. While it does not involve metric functions, information on these functions is very useful for the numerical solution of the constraints.

\section{The dimensionless system of PDEs}\label{App:Dimensionless_system}

For the CDM--baryon problem where both species are assumed to be strictly dust fluids ($p\equiv0$), Eqs.~(\ref{eqn:NrAN}) and~(\ref{eqn:Apr-rhoP}) imply that,
\begin{equation}
A\equiv0, \qquad \hbox{and} \qquad N=1,
\end{equation}
with $A$ defined in (\ref{Hdotu}). Then, redefining the ``dot'' derivative:
\begin{equation}
\hat{\Phi}=\frac{\dot{\Phi}}{H_\ast}=\frac{1}{H_\ast}u^\mu \bigtriangledown_\mu\Phi=\frac{\Phi_{,t}}{H_\ast N}=\frac{\Phi_{,t}}{H_\ast},
\end{equation}
where $H_\ast$ is a constant with inverse--length units sets equal to the initial
background Hubble constant. 

We introduce the following dimensionless parameters and functions:
\begin{eqnarray}
Y=l_\ast \mathcal{Y}, \quad r=l_\ast \xi, \quad \alpha=1/(H_\ast l_\ast),
\\
%
%
%
\mathcal{S}=\frac{\Sigma}{H_\ast},\quad
\mathcal{H}=\frac{H}{H_\ast}, \quad
 \mathcal{W}=\frac{W}{H_\ast^2},\quad
 \chi=\mathcal{Y}_{,\xi},
\\
\mu=\frac{\kappa \rho}{3 H_\ast^2}, \quad \p=\frac{\kappa p}{3 H_\ast^2}, 
\\
\mathcal{M}=\frac{\kappa M}{3 H_\ast^2}, 
\quad
\mathcal{Q}=\frac{\kappa Q}{3 H_\ast^2}, \mathcal{P}=\frac{\kappa \Pi}{3 H_\ast^2},
\end{eqnarray}
with the characteristic length $l_\ast\sim 60$ Mpc. From substituting $0\rightarrow\hbox{DM}$, $i\rightarrow\hbox{B}$, and the above--defined dimensionless
functions  in the system~(\ref{1plus3EFE}), we obtain the desired dimensionless system of PDEs governing the dynamics of a 2--dust--fluid mixture: 
\begin{subequations}\label{2DustSysEqns}
\begin{eqnarray}
\mathcal{\hat  Y} =
 \mathcal{Y}\left(\mathcal{H}+\mathcal{S}\right),
\\
\hat{\chi} = -2\chi \mathcal{S} +\chi \mathcal{H}+\frac{3}{2\alpha} \mathcal{Q} \mathcal{Y},
\\
\hat B = B \left(\mathcal{H}-2 \mathcal{S}\right),
\\
\mathcal{\hat  H}= - \mathcal{H}^2 -2\mathcal{S}^2 - \frac{1}{2} \left(\mu+3 \p\right),  
\\
\mathcal{\hat  S}=\mathcal{S}^2- 2 \mathcal{H} \mathcal{S}+\frac{3}{2} \mathcal{P} - \mathcal{W},
\\
\hat \mu_{\DM} = -3 \mu_{\DM} \mathcal{H},
\\
\hat \mu_{\B} =- 3(\mu_{\B}+p)\mathcal{H}-6 \mathcal{P} \mathcal{S}-\frac{2 \alpha \mathcal{Q} \chi}{\mathcal{Y} B^2}
\nonumber
\\
-\frac{\alpha\mathcal{Q}_{,\xi} }{B^2}+
\frac{\alpha\mathcal{Q} B_{,\xi}}{B^3},
\\
 \mathcal{\hat Q}=-3 \mathcal{H} \mathcal{Q}-\alpha \p_{,\xi}+2\alpha\mathcal{P}_{,\xi}+ \frac{6\alpha\mathcal{P} \chi}{\mathcal{Y}},
\end{eqnarray}
where
\begin{eqnarray}
\mu_{\B}=\gamma^2 \mu^\ast_{\B}, 
\quad
\p\equiv \p_{\B}=\frac{1}{3} \gamma^2  v^2  \mu^\ast_{\B}, 
\\
\mathcal{Q}\equiv \mathcal{Q}_{\B} = \gamma^2 \mu^\ast_{\B} V, 
\quad
\mathcal{P}\equiv \mathcal{P}_{\B}=\frac{1}{3}\gamma^2 \mu^\ast_{\B} v^2,
\\
\mu=\mu_{\DM} + \mu_{\B}.
\end{eqnarray}
At each time the velocity and intrinsic density of the second (non--comoving) dust is determined by:
\begin{equation}
V=\frac{\mathcal{Q}}{\mu_{\B}}, \quad \hbox{and}  \quad \mu_{\B}^\ast =\frac{ \mu_{\B}^2-\mathcal{Q}_{\B}^2 }{\mu_{\B}}.
\end{equation}
The system is complemented by the following constraints
\begin{eqnarray}
\mathcal{H}^2 &=& \mu-k+\mathcal{S}^2, \quad \hbox{with} \quad k=\frac{\mathcal{K}}{H_\ast^2}, 
\\
\mathcal{W}& =& -\frac{\mu}{2}+\frac{\mathcal{M}}{\mathcal{Y}^3}-\frac{3\mathcal{P}}{2},
\end{eqnarray}
where,
\begin{equation}
\mathcal{M}=\frac{1}{2} \mathcal{Y} \left(\mathcal{Y}^2 (\mathcal{S}+\mathcal{H})^2-\frac{\alpha^2\chi^2}{B^2}+1\right).
\end{equation}
\end{subequations}
On the other hand, the equation~(\ref{dotW}) (redundant) results,
\begin{eqnarray}
 \mathcal{\hat W}&+& \frac{3}{2}  \mathcal{\hat P}=-3 \left(\mathcal{H} +\mathcal{S}\right)\mathcal{W} 
 -\frac{3}{2} \left( \mu+\p\right) \mathcal{S}
\nonumber
\\
&{-}&\frac{3}{2} \left(\mathcal{H}-\mathcal{S}\right) \mathcal{P}
- \frac{\alpha}{2} \frac{B_{,\xi} \mathcal{Q}}{B^3}
+\frac{\alpha}{2} \frac{\mathcal{Q}_{,\xi}}{ B^2}-
\frac{\alpha}{2} \frac{\chi \mathcal{Q} }{B^2 \mathcal{Y}},\nonumber\\\label{dimdotW}
\end{eqnarray}
We employed the Method of Lines to integrate this system of PDEs. 
Proceeding in this way the PDEs were discretized along the radial variable, setting
$1000$ grid points within the interval $r/l_\ast\in\left[0,0.2\right]$. The resulting set of ordinary differential equations was 
integrated using an adaptive step--size Runge--Kutta of $4(5)$--th order.


\section{LTB limit}\label{App:LTB_limit}
The LTB model is a general inhomogeneous
spherically symmetric solution of the Einstein's equations for a single irrotational dust fluid as source $T^{\mu \nu}=\rho u^\mu u^\nu$.
The time--synchronous metric can be cast as follows \cite{sussman2013weighed},
\begin{equation}
ds^2=-dt^2+\frac{R^{2}_{,r}(r,t)}{1+2 E(r)} dr^2 + R^2(r,t) \left(d\theta^2+\sin^2(\theta) d\phi^2\right).
\end{equation}
For a comoving 4--velocity $u^\mu =\delta^\mu_{\,t}$ the field equations reduce to:
\begin{equation}
\dot{R}^2=\frac{2 M}{R}+2 E,
\qquad
\hbox{and}
\qquad
M_{,r}=\frac{\kappa}{2} \rho R^2 R_{,r},
\end{equation}
%
Following a similar approach to that used in the main text, we rewrite the Einstein's equations in terms of  covariant objects associated with 
the 4--velocity and the  energy--momentum and projection tensors, which leads to,
\begin{subequations}
\begin{eqnarray}
\dot H &=&-H^2-\frac{\kappa}{6}\,\rho-2\Sigma^2,\label{EV1}
\\
\dot \rho &=& -3\,\rho\,H,\label{EV2}\\
\dot\Sigma &=& -2\,H\,\Sigma+\Sigma^2-\mathcal{W},\label{EV3}
\\
 \dot{\mathcal{W}}  &=& -\frac{\kappa}{2}\,\rho\Sigma-3\,\mathcal{W} \left(H+\Sigma\right),\label{EV4}
\end{eqnarray}
together with the constraint (among others not listed here)
\begin{eqnarray}
H^2 = \frac{\kappa}{3}\, \rho-\mathcal{K}+\Sigma^2,
\label{constr} 
\end{eqnarray}
\end{subequations}
where the expansion scalar, the eigenvalues of the shear and magnetic Weyl tensors as well as the spatial curvature take the simple form: 
\begin{eqnarray}
3 H=\frac{2\dot R}{R}+\frac{\dot R'}{R'}, 
\qquad 
\Sigma =-\frac{1}{3}\left(\frac{\dot R_{,r}}{R_{,r}}-\frac{\dot R}{R}\right),
\\
\mathcal{W}=-\frac{M}{R^3}+\frac{\kappa}{6} \rho,
\qquad 
\mathcal{K}= -\frac{4\left(ER\right)_{,r}}{6 R^2 R_{,r}}. 
\end{eqnarray}
These previous equations are recovered by specializing the system \eqref{1plus3EFE}--\eqref{Mdef} (or its particular case (\ref{2DustSysEqns})) 
to the single comoving fluid case. This can be checked by making the following substitutions:
\begin{equation}
v\rightarrow0, \quad N\rightarrow1, \quad Y\rightarrow R, \quad B\rightarrow \frac{R^\prime}{1+2 E}, \quad 2 E \rightarrow -K.
\end{equation}
%


\providecommand{\noopsort}[1]{}\providecommand{\singleletter}[1]{#1}%

\end{document}